\documentclass[11pt]{article}
\usepackage{amsfonts,amsmath,amssymb}
\usepackage{enumerate}
\usepackage{hyperref}
\usepackage{bbm}
\usepackage{nicefrac}
\usepackage[all]{xy}
\usepackage{graphicx}
\usepackage{bm}
\usepackage{makecell}
\usepackage[table]{xcolor}

\usepackage{upgreek}

\usepackage{float}			
\usepackage{lscape}                 

\usepackage{booktabs}

\newcommand{\be}{\begin{equation}}
\newcommand{\ee}{\end{equation}}

\newcommand{\bea}{\begin{eqnarray}}
\newcommand{\eea}{\end{eqnarray}}

\newcommand{\bes}{\begin{subequations}}
\newcommand{\ees}{\end{subequations}}

\newcommand{\cN}{{\cal N}}
\newcommand{\cF}{{\cal F}}
\newcommand{\cR}{{\cal R}}

\def\sst#1{{\scriptscriptstyle #1}}

\def\0{{\sst{(0)}}}
\def\1{{\sst{(1)}}}
\def\2{{\sst{(2)}}}
\def\3{{\sst{(3)}}}
\def\4{{\sst{(4)}}}
\def\5{{\sst{(5)}}}
\def\6{{\sst{(6)}}}
\def\7{{\sst{(7)}}}
\def\8{{\sst{(8)}}}

\def\cF{{{\cal F}}}

\allowdisplaybreaks  

\usepackage{multirow}
\usepackage{rotating}

\newcommand{\ba}{\begin{align}}
\newcommand{\ea}{\end{align}}

\newcommand{\bse}{\begin{subequations}}
\newcommand{\ese}{\end{subequations}}

\allowdisplaybreaks

\usepackage{lipsum}

\newlength\Colsep
\begin{document}

\makeatletter
\renewcommand{\theequation}{\thesection.\arabic{equation}}
\@addtoreset{equation}{section}
\makeatother

\begin{titlepage}

\begin{flushright}
%
%
\end{flushright}

\vspace{5pt}

   \begin{center}
   \baselineskip=16pt

   \begin{Large}\textbf{
\hspace{-18pt} Super-Chern-Simons spectra \\[8pt]
from Exceptional Field Theory
}
   \end{Large}

\vspace{25pt}

{\large  Oscar Varela }

\vspace{30pt}

	\begin{small}

   {\it Department of Physics, Utah State University, Logan, UT 84322, USA}

	\vspace{15pt}
          
   {\it Departamento de F\'\i sica Te\'orica and Instituto de F\'\i sica Te\'orica UAM/CSIC , \\
   Universidad Aut\'onoma de Madrid, Cantoblanco, 28049 Madrid, Spain} 
		
	\end{small}

\vskip 70pt

\end{center}

\begin{center}
\textbf{Abstract}
\end{center}

\begin{quote}

Exceptional Field Theory has been recently shown to be very powerful to compute Kaluza-Klein spectra. Using these techniques, the mass matrix of Kaluza-Klein vector perturbations about a specific class of AdS$_4$ solutions of $D=11$ and massive type IIA supergravity is determined. These results are then employed to characterise the complete supersymmetric spectrum about some notable $\cN=2$ and $\cN=3$ AdS$_4$ solutions in this class, which are dual to specific three-dimensional superconformal Chern-Simons field theories. 

\end{quote}

\vfill

\end{titlepage}

\tableofcontents



\section{Introduction}


For holographic conformal field theories (CFTs), the spectrum of single-trace operators with scaling dimension of order one, at strong coupling and large $N$, is mapped to the spectrum of Kaluza-Klein (KK) perturbations about their dual anti-de Sitter (AdS) solutions of string or M-theory \cite{Maldacena:1997re,Gubser:1998bc,Witten:1998qj}. The KK spectrum of Type IIB supergravity on the background AdS$_5 \times S^5$ \cite{Schwarz:1983qr}, relevant to  $\cN=4$ super-Yang-Mills, was computed in \cite{Kim:1985ez}. The face-value calculation of \cite{Kim:1985ez} entails complicated field redefinitions, a demanding linearisation of the type IIB equations of motion, an involved expansion of the linearised fields in scalar, spinor and vector spherical $S^5$ harmonics and, finally, a diagonalisation of the resulting mass matrices. In this particular case with maximal (super)symmetry, all fields turn out to fill out short supermultiplets of the four-dimensional maximally supersymmetric conformal algebra. For this reason, both the algebraic structure of the spectrum and the physical masses of all fields are actually dictated by group theory \cite{Gunaydin:1984fk}. A similar remark applies to the KK spectrum \cite{Englert:1983rn,Sezgin:1983ik,Biran:1983iy} of the maximally (super)symmetric $D=11$ Freund-Rubin solution AdS$_4 \times S^7$ \cite{Freund:1980xh}, dual to ABJM \cite{Aharony:2008ug}. 

For AdS/CFT dual pairs with less (super)symmetry, group theory still determines the algebraic structure of the spectrum, namely, the possible supermultiplets in given representations of the residual symmetry group that are present in the spectrum. Typically, these spectra still contain short multiplets, whose conformal dimensions are again fixed by representation theory. But, unlike in the maximally supersymmetric cases, long multiplets will usually be contained in the spectra as well. For these multiplets, group theory only requires that a unitarity bound be respected but, other than this, has no power to predict the actual value of their dimensions. Thus, for AdS solutions with less than maximal supersymmetry, there is no alternative to computing the long KK spectrum other than direct calculation. For homogeneous Freund-Rubin-type solutions, this problem can be attacked using coset space technology \cite{Fabbri:1999mk,Ceresole:1999ht,Fre:1999gok,Nilsson:2018lof}. More generally, though, coset methods will not be available for AdS solutions with inhomogeneous internal geometries of relatively small symmetry, supported by fluxes and warp factors. If the direct calculation of the spectrum \cite{Kim:1985ez} of the maximally supersymmetric background AdS$_5 \times S^5$ \cite{Schwarz:1983qr} was so demanding, the calculation of KK spectra about inhomogeneous AdS solutions with less (super)symmetry is downright prohibitive with the traditional methods of \cite{Kim:1985ez}. 

Recently, a powerful alternative to the techniques of \cite{Kim:1985ez} for the calculation of KK spectra about certain AdS solutions of string or M-theory has been put forward \cite{Malek:2019eaz,Malek:2020yue} based on Exceptional Field Theories (ExFTs) \cite{Hohm:2013pua,Hohm:2013vpa,Hohm:2013uia} (see \cite{Berman:2020tqn} for a review). Like their Exceptional Generalised Geometry cousins \cite{Pacheco:2008ps,Berman:2010is}, these correspond to reformulations of the $D=10$ and $D=11$ supergravities where the exceptional symmetries of the lower-dimensional maximal supergravities, {\it e.g.}~E$_{7(7)}$ for $D=4$ which is fixed henceforth, are explicitly realised. This is done at the expense of reducing the manifest $D=10$ or $D=11$ local Lorentz covariance to only a manifest $D=4$ local Lorentz covariance. Even if, due to the latter feature, the ExFTs superficially look four-dimensional, they still are fully-fledged higher-dimensional theories. For this reason, not only the AdS solutions of the higher-dimensional theories can be recovered as solutions of ExFT, but also the full towers of KK perturbations about these AdS solutions are contained within ExFT as well. 

The reason why ExFT methods to compute KK spectra \cite{Malek:2019eaz,Malek:2020yue} have an edge over the traditional approach \cite{Kim:1985ez} is essentially two-fold. On the one hand, the complicated field redefinitions needed prior to linearisation are already built-in (at the full non-linear level, in fact) into ExFT. On the other hand, all fields need to be expanded only in {\it scalar} harmonics of the ordinary internal manifold. Unlike the method of \cite{Kim:1985ez}, though,  the ExFT technology \cite{Malek:2019eaz,Malek:2020yue} comes with a regime of applicability which hinges also on two assumptions. Namely, that the relevant AdS solutions have an associated consistent truncation to lower-dimensional maximal gauged supergravity, and that the internal space in the ordinary $D=10$ or $D=11$ sense, must be topologically spherical. These are not severe limitations, since the class of AdS solutions of this type still comes in cornucopious  abundance \cite{Bobev:2020ttg,Krishnan:2020sfg,Comsa:2019rcz,Guarino:2020jwv}. Furthermore, having fixed an allowed lower-dimensional gauging, the same set of spherical harmonics on the associated round sphere is valid to compute the KK spectra about {\it any} other AdS solution that uplifts from the same gauging, even if the round sphere is not a (supersymmetric) solution itself.

Within their validity regime, these ExFT techniques \cite{Malek:2019eaz} certainly outperform the standard methods of \cite{Kim:1985ez}. One must nevertheless linearise the field equations of ExFT which, for certain fields, particularly the internal scalar fields, is still a rather involved task. Unlike the scalars', the linearisation of the ExFT vector and graviton equations of motion is significantly more manageable. Fortunately, for AdS solutions with sufficiently high supersymmetry, $\cN \geq 2$ in $D=4$, an explicit calculation of the spectrum of KK scalars (and spin $1/2$ fermions) is not  necessary and can be indirectly inferred from the vector and graviton spectra, and from group theory. The reason, for $\cN=2$, is that the only scalars and fermions that belong to OSp$(4|2)$ multiplets which do not contain gravitons or vectors, must necessarily belong to short hypermultiplets. And, for these, their dimensions are fixed by the R-charges, similarly to the maximally supersymmetric cases discussed above. The situation is even more restrictive for $\cN=3$, for which already the vector multiplets are necessarily short. In these $\cN \geq 2$  cases, the complete supersymmetric KK spectra are thus fixed by group theory together with the graviton and vector spectra. With this in mind, section \ref{sec:KKVecMassMat} presents a derivation of the KK vector mass matrix from ExFT. An alternative, though equivalent, form of the KK vector mass matrix can be found in \cite{Malek:2020yue}. 

In this paper, I will compute the complete supersymmetric KK spectrum about the supersymmetric AdS$_4$ solutions of $D=11$ \cite{Cremmer:1978km} and massive type IIA supergravity \cite{Romans:1985tz} that uplift from $D=4$ $\cN=8$ supergravity with specific gaugings. More concretely, I will focus on the $D=11$ $\cN=2$ AdS$_4$ solution found by Corrado, Pilch and Warner (CPW) \cite{Corrado:2001nv}. For this solution, the OSp$(4|2)$ supermultiplet structure of the spectrum was elucidated in \cite{Klebanov:2008vq}, the graviton spectrum was computed in \cite{Klebanov:2009kp} using specific methods for spin-2 (see \cite{Bachas:2011xa}), and the complete spectrum has been recently given in \cite{Malek:2020yue}. In section \ref{sec:N=2Spectrum11D}, I compute the spectrum of vectors and allocate them in supermultiplets, finding agreement with \cite{Malek:2020yue}. I will also characterise the complete supersymmetric KK spectrum about two specific AdS$_4$ solutions of massive IIA supergravity with $\cN=2$ \cite{Guarino:2015jca} and $\cN=3$ \cite{Pang:2015vna,DeLuca:2018buk} supersymmetry. The graviton spectra for these solutions was respectively computed in \cite{Pang:2017omp} and \cite{Pang:2015rwd}. By the previous arguments, all that is left to determine the complete supersymmetric KK spectra is an analysis of their OSp$(4|2)$ or OSp$(4|3)$ supermultiplet structure and the calculation of the KK vector spectra. These items are addressed in section \ref{sec:SpectraIIA}. Section \ref{sec:Discussion} concludes, and some appendices contain useful supplementary material.

\vspace{20pt}

\section{KK vector mass matrix from ExFT} \label{sec:KKVecMassMat}

The starting point for the present analysis is the E$_{7(7)}$ ExFT \cite{Hohm:2013uia} reformulation of $D=11$ supergravity \cite{Cremmer:1978km}, and its deformation thereof \cite{Ciceri:2016dmd} suitable to accommodate massive type IIA supergravity \cite{Romans:1985tz}. Both theories, \cite{Hohm:2013uia} and \cite{Ciceri:2016dmd}, can be treated simultaneously for the purposes of this analysis, and both of them will be collectively referred to as ExFT. Let ${\cal A}_\mu^M (x, Y)$ be the gauge fields present in the theory, and ${\cal F}^M_{\mu \nu} (x,Y)$ their field strengths. These depend on both the external, $x^\mu$, $\mu = 0 , 1 ,2 ,3 $, and the internal, $Y^M$, $M =1, \ldots, 56$, ExFT coordinates. The curved index $M$ labels the fundamental representation of E$_{7(7)}$, and the gauge field, as well as any other field in the theory, is subject to the relevant section constraints. These will not be explicitly needed. The objective of this section is to linearise the field equations of ${\cal A}_\mu^M (x, Y)$ and find the mass matrix for the KK vectors about the class of AdS$_4$ backgrounds of $D=11$ supergravity and massive IIA supergravity that uplift from the $D=4$ $\cN=8$ gaugings to be specified momentarily.

\newpage 

\subsection{Preliminaries} \label{sec:KKVecPrelim}

The ExFT (pseudo-)Lagrangian involves ${\cal A}_\mu^M$ or ${\cal F}^M_{\mu \nu} $ in four instances \cite{Hohm:2013uia,Ciceri:2016dmd}: the Einstein-Hilbert term for the external metric $\bm{g}_{\mu\nu} (x,Y)$, the topological term for ${\cal F}^M_{\mu \nu} $, the kinetic term for ${\cal F}^M_{\mu \nu} $ and the kinetic term the internal scalar metric ${\cal M}_{M N} (x,Y)$. Neither the first nor the second terms are expected to contribute to the vector mass matrix: the Einstein-Hilbert term contains the couplings of the metric to the vectors and should be partially responsible for Higgsing the KK gravitons. The topological terms for ${\cal F}^M_{\mu \nu} $ in the ExFT (pseudo-)Lagrangian can be eliminated on-shell in favour of the gauge equations of motion and, for that reason, can also be disregarded. Thus, one is led to focus on the gauge kinetic terms and the scalar kinetic terms of the Lagrangian \cite{Hohm:2013uia}:
\begin{equation} \label{eq:ExFTVecs}
{\cal L}_{\textrm{ExFT vectors}} = - \tfrac18 \,\bm{e} \, {\cal M}_{M N} \, {\cal F}^M_{\mu \nu} \, {\cal F}^{N \mu \nu} + \tfrac{1}{48} \,  \bm{e} \, {\cal D}_\mu {\cal M}_{MN} \,  {\cal D}^\mu {\cal M}^{MN} .
\end{equation}
Here, $\bm{e} \equiv \sqrt{ | \textrm{det} \,  \bm{g}_{\mu\nu} | }$, external indices are raised with the inverse metric $\bm{g}^{\mu\nu}$, and ${\cal M}^{MN}$ is the matrix inverse to ${\cal M}_{MN}$. The covariant derivative of ${\cal M}_{MN}$ is \cite{Hohm:2013uia,Ciceri:2016dmd}
\begin{equation} \label{eq:CovDerM}
{\cal D}_\mu {\cal M}_{MN} =
 \partial_\mu {\cal M}_{MN} - {\cal A}_\mu^K \, \partial_K {\cal M}_{MN} -24 \;  \partial_K {\cal A}_\mu^L \; \mathbb{P}^K{}_L{}^P{}_{(M} {\cal M}_{N)P} - 2 \, {\cal A}_\mu^K \, F_{K(M}{}^L \, {\cal M}_{N)L}    \; ,
\end{equation}
and similarly for ${\cal D}_\mu {\cal M}^{MN}$. In (\ref{eq:CovDerM}), the projector onto the adjoint of E$_{7(7)}$ has been introduced as 
\begin{equation} \label{eq:ProjE7}
\mathbb{P}^M{}_N{}^P{}_Q \equiv (t_\alpha)_N{}^M (t^\alpha)_Q{}^P \; ,
\end{equation}
with $(t_\alpha)_N{}^M$ the E$_{7(7)}$ generators. The E$_{7(7)}$ adjoint index $\alpha =1 , \ldots , 133$, is raised in (\ref{eq:ProjE7}) and elsewhere with the inverse, $\kappa^{\alpha \beta}$, of the Cartan-Killing form $\kappa_{\alpha \beta} = (t_\alpha)_M{}^N (t_\beta)_N{}^M$. Finally, $F_{MN}{}^P$ in (\ref{eq:CovDerM}) is a deformation \cite{Ciceri:2016dmd} of the original ExFT formulation of \cite{Hohm:2013uia}. For $D=11$ configurations, $F_{MN}{}^P =0$, and for type IIA, $F_{MN}{}^P$ encodes a magnetic gauging contribution from the Romans mass: see \cite{Ciceri:2016dmd} for the details.

I would like to further restrict the problem to the specific configurations of ExFT that give rise to a $D=4$ ${\cal N} = 8$ gauged supergravity upon consistent truncation. On top of this configuration, $D=4$ KK vector perturbations will also be kept. The concrete $D=4$ $\cN=8$ gaugings that I will consider will be the SO(8) gauging \cite{deWit:1982ig} and the dyonic \cite{Dall'Agata:2012bb,Dall'Agata:2014ita,Inverso:2015viq} ISO(7) gauging \cite{Guarino:2015qaa}. Both of these indeed arise after consistent truncation of $D=11$ and massive IIA supergravity on $S^7$ \cite{deWit:1986iy} and $S^6$ \cite{Guarino:2015jca,Guarino:2015vca}, respectively. Under these assumptions, the external and internal ExFT metrics $\bm{g}_{\mu\nu}$ and ${\cal M}_{MN}$ are set to \cite{Berman:2012uy,Lee:2014mla,Hohm:2014qga,Cassani:2016ncu} 
\begin{equation} \label{eq:ScalarMetExFT}
\bm{g}_{\mu\nu}(x,Y)  = \rho(Y)^{-2} \, g_{\mu\nu} (x) \; , \qquad
{\cal M}_{MN}  (x, Y) = U_{(M}{}^{\underline{M}} (Y) \, U_{N)}{}^{\underline{N}} (Y) \, M_{ \underline{M} \underline{N} } (x)  \; . 
\end{equation}
Here, $g_{\mu\nu}$ and $M_{ \underline{M} \underline{N} }$ respectively correspond to the $D=4$ $\cN=8$ metric and $\textrm{E}_{7(7)}/\textrm{SU}(8)$ scalar matrix. The $Y$-dependent function $\rho$ and the generalised Scherk-Schwarz twist matrix $U_{M}{}^{\underline{N}}$ are needed in (\ref{eq:ScalarMetExFT}) for consistency. Their explicit expressions will not be needed. All that will be necessary to know from them is that they must obey the (generalised parallelisability) relations \cite{Lee:2014mla,Hohm:2014qga,Cassani:2016ncu,Inverso:2017lrz} 
\begin{eqnarray} \label{eq:ConsistentKK}
&& \partial_N \, (U^{-1})_{\underline{M}}{}^N - 3 \rho^{-1}  (U^{-1})_{\underline{M}}{}^N\partial_N \, \rho =0 \; , \nonumber \\[5pt]
&& 7 \rho^{-1} \, \left( (U^{-1})_{\underline{M}}{}^P \, (U^{-1})_{\underline{N}}{}^Q \, \partial_P \, U_Q{}^{\underline{K} }\right)_{\bm{912}} + F_{\underline{M} \underline{N}}{}^{\underline{K}}  = \Theta_{\underline{M}}{}^\alpha (t_\alpha)_{\underline{N}}{}^{\underline{K}} \; ,
\end{eqnarray}
which ensure the consistency of the ExFT truncation to $D=4$ $\cN=8$ gauged supergravity with (constant, $x$- and $Y$-independent) embedding tensor $\Theta_{\underline{M}}{}^\alpha$. In (\ref{eq:ConsistentKK}), $()_{\bm{912}}$ denotes projection to the $\bm{912}$ representation of E$_{7(7)}$. The indices $ \underline{M}= 1 , \ldots , 56$ are flat, fundamental E$_{7(7)}$ indices, as pertains to strictly $D=4$ quantities. Quantities with curved and flat E$_{7(7)}$ indices are related through the twist matrix $U_M{}^{\underline{N} }$ and its inverse.

When the dictionary that relates ExFT to $D=11$ or $D=10$ supergravities \cite{Hohm:2014qga,Cassani:2016ncu}  is employed, the expressions (\ref{eq:ScalarMetExFT}), together with those for the remaining ExFT fields, give rise to the explicit embedding of $D=4$ $\cN=8$ supergravity in the higher-dimensional theories. The scalar potentials of the SO(8) \cite{deWit:1982ig} and ISO(7) \cite{Guarino:2015qaa} $D=4$ gaugings of interest attain AdS$_4$ vacua, and these correspondingly uplift to $D=11, 10$. On top of the $D=4$ $\cN=8$ supergravity fields, I would also like to extract the infinite tower of $D=4$ KK vector perturbations about any of these AdS$_4$ vacua. A useful parametrisation for the infinite tower of KK vector pertubations is $A_\mu^{\underline{M} \Lambda} (x) $ \cite{Malek:2019eaz,Malek:2020yue}, in terms of a combined direct-product index $\underline{M} \Lambda$. Here, $\underline{M}$ is a flat index in the $\bm{56}$ of E$_{7(7)}$, and the index
\begin{equation} \label{eq:SigmaIndex}
\Lambda =  \big( 1, A_1  \, , \,  \{A_1 A_2 \}  \, , \, \ldots ,  \{A_1  \ldots A_n  \} \, , \, \ldots  \big)  \; \textrm{or} \;  
\Lambda =  \big( 1, I_1  \, , \,  \{I_1 I_2 \}  \, , \, \ldots ,  \{I_1 \ldots I_n  \} \, , \, \ldots  \big) \; , 
\end{equation}
ranges on the infinite-dimensional, reducible representations 
\begin{equation} \label{eq:SymTrac}
\oplus_{n=0}^\infty [n,0,0,0] \; \textrm{of SO(8)} \quad \textrm{or} \quad   \oplus_{k=0}^\infty  [k,0,0] \; \textrm{of SO(7)} \; ,
\end{equation}
for AdS$_4$ solutions that uplift from the $D=4$ SO(8) or ISO(7) gaugings, respectively, regardless or their actual residual symmetry $G \subset \textrm{SO}(8)$ or $G \subset \textrm{SO}(7)$. The curly brackets in (\ref{eq:SigmaIndex}) denote traceless symmetrisation, while $A_1 = 1 , \ldots , 8$ and $I_1 = 1 , \ldots , 7$ here and elsewhere denote fundamental indices of SO(8) (in the $\bm{8}_v$) and SO(7) (in the $\bm{7}$). 

The KK gauge fields $A_\mu^{\underline{M} \Lambda} (x) $ and their linearised field strengths $F^{\underline{M}\Sigma}_{\mu \nu} (x) = 2 \,   \partial_{ [ \mu } A^{\underline{M}\Sigma}_{\nu]} (x) $ are embedded into their ExFT counterparts ${\cal A}_\mu^{M} (x,Y) $ and ${\cal F}^{M}_{\mu \nu} (x,Y) = 2 \,   \partial_{ [ \mu } {\cal A}^{M}_{\nu]} (x,Y) +\ldots $ (the latter linearised as well for the present purposes) through \cite{Malek:2019eaz,Malek:2020yue}
\begin{equation} \label{eq:ExFTVecsKK}
{\cal A}^M_\mu = \rho^{-1} (U^{-1})_{\underline{N}}{}^M A_\mu^{\underline{N} \Lambda} \, {\cal Y}_\Lambda
 \; , \qquad 
{\cal F}^M_{\mu \nu} = \rho^{-1} (U^{-1})_{\underline{N}}{}^M F_{\mu \nu} ^{\underline{N} \Lambda} \, {\cal Y}_\Lambda \; .
\end{equation}
Here, ${\cal Y}_\Lambda$ denotes the infinite tower of scalar spherical harmonics on the round $S^7$ or $S^6$ spheres. These lie in the representations of SO(8) or SO(7) indicated in (\ref{eq:SymTrac}). The explicit expressions of ${\cal Y}_\Lambda$ in terms of $S^7$ or $S^6$ coordinates will not be needed: suffice it to note that the action of the Scherk-Schwarz twist matrix on these is given by \cite{Malek:2019eaz,Malek:2020yue}
\begin{equation} \label{eq:ActionSH}
\rho^{-1} \, (U^{-1})_{\underline{N}}{}^M \partial_M \, {\cal Y}_\Lambda = -({\cal T}_{\underline{N}})_\Lambda{}^\Sigma \, {\cal Y}_\Sigma \; .
\end{equation}
The (constant, $x$- and $Y$-independent) matrices $({\cal T}_{\underline{N}})_\Lambda{}^\Sigma$ correspond to the generators of SO(8) or SO(7) in the infinite-dimensional, reducible representations (\ref{eq:SymTrac}), normalised as 
\begin{equation}
[ {\cal T}_{\underline{M}} , {\cal T}_{\underline{N}} ] = -X_{\underline{M} \underline{N}}{}^{\underline{P}} \, {\cal T}_{\underline{P}} \; , \qquad \textrm{with} \quad
X_{\underline{M} \underline{N}}{}^{\underline{P}} \equiv \Theta_{\underline{M}}{}^\alpha \, (t_\alpha)_{\underline{N}}{}^{\underline{P}} \; .
\end{equation}

\newpage

\subsection{The KK vector mass matrix} \label{sec:KKVecMassMatCalc}

Equipped with the definitions introduced in section \ref{sec:KKVecPrelim}, the goal is now to linearise the ExFT vector equations of motion or, equivalently, to retain the quadratic terms in the action (\ref{eq:ExFTVecs}), in order to read off the mass matrix for the KK vectors $A_\mu^{\underline{M} \Lambda} $. 

The action (\ref{eq:ExFTVecs})  is already manifestly quadratic in the ExFT gauge fields ${\cal A}_\mu^{M}  $ and in their linearised field strengths ${\cal F}^{M}_{\mu \nu} (x) = 2 \,   \partial_{ [ \mu } {\cal A}^{M}_{\nu]} (x,Y) +\ldots $. 
Inserting (\ref{eq:ScalarMetExFT}) and the left-most relation in (\ref{eq:ExFTVecsKK}) into the covariant derivative (\ref{eq:CovDerM}), some calculation allows one to compute 
\begin{equation} \label{eq:CovDerMM}
{\cal D}_\mu {\cal M}_{MN} = U_{(M}{}^{\underline{M}} \, U_{N)}{}^{\underline{N}}
\Big[ \partial_\mu M_{ \underline{M} \underline{N} } - 2 \, A_\mu^{\underline{K} \Sigma}  \left( \Theta_{\underline{K}}{}^\alpha \, \delta_\Sigma^\Lambda -12 \, (t^\alpha)_{\underline{K}}{}^{\underline{L}} \, ({\cal T}_{\underline{L}})_\Sigma{}^\Lambda \right)  (t_\alpha)_{(\underline{M}}{}^{\underline{P}} M_{\underline{N}) \underline{P}} {\cal Y}_\Lambda \Big]  , 
\end{equation}
and similarly for ${\cal D}_\mu {\cal M}^{MN}$. In order to reach this result, the consistent truncation conditions (\ref{eq:ConsistentKK}) and the action (\ref{eq:ActionSH}) on the spherical harmonics need to be used. The explicit form for the projector to the $\bm{912}$ representation of E$_{7(7)}$, which can be found in {\it e.g.}~\cite{deWit:2002vt}, is also needed. All the dependences on the internal coordinates $Y$ brought by $\rho$ and $U_M{}^{\underline{N} }$ through (\ref{eq:ScalarMetExFT}), (\ref{eq:ExFTVecsKK}), (\ref{eq:CovDerMM}) drop out at the level of the ExFT field equations \cite{Hohm:2014qga}. The only dependences on the internal ExFT coordinates are brought into the Lagrangian (\ref{eq:ExFTVecs}) through a quadratic combination ${\cal Y}_\Lambda \, {\cal Y}_\Sigma$  of spherical harmonics. Under the integral sign at the level of the action, this dependence simply becomes $\delta_{\Lambda \Sigma}$ by virtue of the orthogonality of the spherical harmonics. 

Putting all these contributions together, and considering similar contributions from the kinetic terms for ${\cal F}^{M}_{\mu \nu}$, the action (\ref{eq:ExFTVecs}) yields
\begin{equation} \label{eq:ExFTVecsKKLag}
{\cal L}_{\textrm{KK vectors}} = - \tfrac18 \, e \, M_{\underline{M} \underline{N}} \, \delta_{\Lambda \Sigma} \, F^{\underline{M}\Lambda}_{\mu \nu} \, F^{ \mu \nu \underline{N} \Sigma} + \tfrac14  \, e \, ({\cal N}^2)_{\underline{M} \Lambda , \underline{N} \Sigma }  \, A^{\underline{M}\Lambda}_{\mu } \, A^{ \mu \underline{N} \Sigma } \; .
\end{equation} 
Here, $e \equiv \sqrt{ | \textrm{det} \,  g_{\mu\nu} | }$, external indices are raised with the $D=4$ inverse metric $g^{\mu\nu}$, and $({\cal N}^2)_{\underline{M} \Lambda , \underline{N} \Sigma } $ is the symmetrisation in the combined index $\underline{M} \Lambda$,
\begin{equation} \label{eq:MassAux1}
({\cal N}^2)_{\underline{M} \Lambda , \underline{N} \Sigma } \equiv \tfrac12 \, (\widetilde{\cal N}^2)_{\underline{(M |} (\Lambda |, \underline{|N)} |\Sigma ) } +\tfrac12 \,  (\widetilde{\cal N}^2)_{\underline{[M |} [\Lambda |, \underline{|N]} |\Sigma ] }
\end{equation}
of the quantity
{\setlength\arraycolsep{2pt} 
\begin{eqnarray} \label{eq:MassAux2}
(\widetilde{\cal N}^2)_{\underline{M} \Lambda , \underline{N} \Sigma }  \equiv &&
\tfrac{1}{3} \,  \delta_{\Omega\Omega^\prime} \,  \Big( \Theta_{\underline{M}}{}^\alpha \, \delta_\Lambda^\Omega -12 \, (t^\alpha)_{\underline{M}}{}^{\underline{P}} \, ({\cal T}_{\underline{P}})_\Lambda{}^\Omega \Big)
\Big( \Theta_{\underline{N}}{}^\beta \, \delta_\Sigma^{\Omega^\prime} -12 \, (t^\beta)_{\underline{N}}{}^{\underline{Q}} \, ({\cal T}_{\underline{Q}})_\Sigma{}^{\Omega^\prime} \Big) \nonumber \\[5pt]
&& \qquad  \quad \times  (t_\alpha)_{(\underline{R}}{}^{\underline{T}} \, M_{\underline{S}) \underline{T} } \,  (t_\beta)_{\underline{U}}{}^{(\underline{R}} \, M^{\underline{S}) \underline{U} } \; .
\end{eqnarray}
}The matrix $M^{\underline{M} \underline{N}}$ here is the inverse of the $D=4$ $\cN=8$ scalar matrix $M_{\underline{M}\underline{N}}$.

The KK vector mass matrix $({\cal M}^2)_{\underline{M} \Lambda}{}^{\underline{N} \Sigma} $ is finally obtained by canonically normalising the kinetic term in (\ref{eq:ExFTVecsKKLag}), as usual. The result is thus
\begin{equation} \label{eq:KKVecMassMat}
({\cal M}^2)_{\underline{M} \Lambda}{}^{\underline{N} \Sigma} = ({\cal N}^2)_{\underline{M} \Lambda , \underline{P} \Omega } \, M^{\underline{P} \underline{N}} \, \delta^{\Omega \Sigma} \; ,
\end{equation}
with $({\cal N}^2)_{\underline{M} \Lambda , \underline{P} \Omega } \, M^{\underline{P} \underline{N}} $ given by (\ref{eq:MassAux1}), (\ref{eq:MassAux2}). A KK vector mass matrix for the AdS$_5$ solutions of type IIB that consistently uplift on $S^5$ \cite{Lee:2014mla,Ciceri:2014wya,Baguet:2015sma} from the $D=5$ $\cN=8$ SO(6) gauging \cite{Gunaydin:1984qu} was derived from E$_{6(6)}$ ExFT \cite{Hohm:2013vpa} in \cite{Malek:2019eaz}. Equation (\ref{eq:KKVecMassMat}) extends that result to the KK vector mass matrix for $D=11$ or massive IIA AdS$_4$ solutions that respectively uplift on $S^7$ \cite{deWit:1986iy} or $S^6$  \cite{Guarino:2015jca,Guarino:2015vca} from the SO(8) \cite{deWit:1982ig} or dyonic ISO(7) \cite{Guarino:2015qaa} gaugings of $D=4$ $\cN=8$ supergravity. An alternative form of the $D=4$ KK vector mass matrix (\ref{eq:KKVecMassMat}) is given in \cite{Malek:2020yue}.

The infinite-dimensional KK vector mass matrix $({\cal M}^2)_{\underline{M} \Lambda}{}^{\underline{N} \Sigma} $ in (\ref{eq:KKVecMassMat}) is block diagonal, with blocks of dimension $( 56 \cdot \textrm{dim} \, [n,0,0,0] ) \times ( 56 \cdot \textrm{dim} \, [n,0,0,0] ) $ or $( 56 \cdot \textrm{dim} \, [k,0,0] ) \times ( 56 \cdot \textrm{dim} \, [k,0,0] ) $ at SO(8) KK level $n=0,1, \ldots$ or SO(7) KK level $k=0,1, \ldots$. At zero-th KK level, the contributions from the generators ${\cal T}_{\underline{M}}$ must be disregarded, and (\ref{eq:KKVecMassMat}) reduces to the vector mass matrix of $D=4$ $\cN=8$ gauged supergravity: see {\it e.g.}~(4.85) of \cite{Trigiante:2016mnt}. More generally, the fact that (\ref{eq:KKVecMassMat}) is block diagonal implies that it can be conveniently diagonalised KK level by KK level, something that does not always happen for the KK mass matrices of other AdS$_4$ solutions outside of the class we are considering \cite{Nilsson:2018lof,Cesaro:2020piw}. It is thus enough to diagonalise each separate block individually and, for KK levels $n \geq 1$, $k \geq 1$, consider the (electric) SO(8) or SO(7) generators, ${\cal T}_{\underline{M} } = \left( {\cal T}_{AB} \, , \,  {\cal T}^{AB} \equiv 0  \right)$ in the symmetric traceless representation $[n,0,0,0]$ or $[k,0,0]$: see appendix \ref{sec:SpecificSpectra} for further details. 

The mass matrix (\ref{eq:KKVecMassMat}) contains spurious eigenvalues that must be removed from the physical spectrum. At SO(8) KK level $n=0 , 1 , 2, \ldots $ (and similarly for SO(7)), $({\cal M}^2)_{\underline{M} \Lambda}{}^{\underline{N} \Sigma} $ always has $28 \cdot \textrm{dim} \, [n,0,0,0]$ unphysical zero eigenvalues, corresponding to the dual, magnetic KK vectors. The electric vector eigenvalues that Higgs the KK gravitons are unphysical as well. For the $D=11$ solutions, the $28 \cdot \textrm{dim} \, [n,0,0,0]$ electric vectors at level $n$ come in the SO(8) representations
\begin{equation} \label{eq:28xSymTrSO8}
[0,1,0,0] \times [n,0,0,0] \, \longrightarrow \, \underline{[n,0,0,0] } + [n,1,0,0] + [n-1,0,1,1] + [n-2,1,0,0] \; , 
\end{equation}
or their branchings under $G \subset \textrm{SO}(8)$ for AdS$_4$ solutions with residual symmetry group $G$. 
In the massive IIA case, the $(21 +7) \cdot \textrm{dim} \, [k,0,0]$ electric vector eigenvalues come instead in the SO(7) representations
\begin{eqnarray} \label{eq:28xSymTrSO7}
\left( \,  [0,1,0] + [1,0,0] \, \right) \times [k,0,0] & \longrightarrow & \underline{[k,0,0] } + [k,1,0] + [k-1,0,2]+ [k-2,1,0] \nonumber  \\
&& +[k+1,0,0] + [k-1,1,0] + [k-1,0,0]  \; ,
\end{eqnarray}
possibly branched out again under $G \subset \textrm{SO}(7)$ for solutions with symmetry $G$. The representations with negative Dynkin labels must be disregarded in (\ref{eq:28xSymTrSO8}), (\ref{eq:28xSymTrSO7}), as they are actually not present. The underlined symmetric traceless representations are not present either for $n=k=0$ and, for $n >1$, $k>1$, must still be disregarded as they contain the unphysical eigenvalues corresponding to the vectors eaten (together with additional scalars) by the $[n,0,0,0]$ or $[k,0,0]$ massive KK gravitons. These unphysical eigenvalues are not typically zero.

Only the eigenvalues of the mass matrix (\ref{eq:KKVecMassMat}) that come in the representations not underlined in (\ref{eq:28xSymTrSO8}), (\ref{eq:28xSymTrSO7}), or their branchings thereof under $G \subset \textrm{SO}(8)$ or $G \subset \textrm{SO}(7)$ for solutions with symmetry $G$, furnish the physical spectrum of KK vectors. The only physical zero eigenvalues of $({\cal M}^2)_{\underline{M} \Lambda}{}^{\underline{N} \Sigma} $ occur at KK level $n=0$ with degeneracy $\dim \, G$, corresponding to the massless KK vectors in the adjoint of $G \subset \textrm{SO}(8)$, and similarly for the ISO(7) gauging. All other vectors at KK level zero and higher are massive. See appendix \ref{sec:SpecificSpectra} for further details on the mass matrix (\ref{eq:KKVecMassMat}), and for the KK vector spectra of selected AdS$_4$ solutions of $D=11$ and massive type IIA supergravity.


\section{The complete KK spectrum of the $\cN=2$ CPW solution } \label{sec:N=2Spectrum11D}


Four-dimensional $\cN=8$ SO(8) gauged supergravity has an $\cN=2$, $\textrm{SU}(3) \times \textrm{U}(1)$-invariant critical point \cite{Warner:1983vz}, which uplifts on $S^7$ to the CPW  AdS$_4$ solution of $D=11$ supergravity with the same (super)symmetry \cite{Corrado:2001nv}. The complete spectrum of the CPW solution is now known and, for that reason, this presentation will be brief. A more detailed analysis for the analogue $\cN=2$ solution of massive type IIA \cite{Guarino:2015jca} will be discussed in section \ref{sec:SpectraIIA}.

The KK spectra for all fields at KK level $n=0$ has long been known \cite{Nicolai:1985hs}, due to the fact that this level agrees, when the non-linear interactions within it are restored, with $D=4$ $\cN=8$ SO(8) supergravity. More recently, the KK level $n=1$ spectrum was computed in \cite{Malek:2019eaz} using ExFT techniques, and extended to higher levels in \cite{Malek:2020yue}. The full KK graviton spectrum is known \cite{Klebanov:2009kp}, as is the generic structure of the entire KK spectrum in representations of $\textrm{OSp}(4|2) \times \textrm{SU}(3)$ \cite{Klebanov:2008vq}. Here, I recover the complete spectrum of this solution \cite{Klebanov:2008vq,Malek:2020yue} by putting together the group theory analysis of \cite{Klebanov:2008vq}, the graviton spectrum of \cite{Klebanov:2009kp}, and the present calculation of the vector spectrum.

The KK vector spectrum for the CPW solution is presented for the first three KK levels, $n=0,1,2$, in the entry labelled as $\cN=2$ U(3) in table \ref{tab:KKVectorsSO8} of appendix \ref{sec:SpecificSpectra}. Firstly, the multiplicities shown in the table are compatible with the $\textrm{OSp}(4|2) \times \textrm{SU}(3)$  group theory of \cite{Klebanov:2008vq}. Secondly, all individual vectors that enter short graviton, short gravitino and short vector multiplets do indeed have their masses fixed by the conformal dimension of those multiplets as given in \cite{Klebanov:2008vq}. Thirdly, there exist vector masses in the tables compatible with those predicted by the KK graviton analysis of \cite{Klebanov:2009kp}. The remaining individual vectors must arrange themselves in long gravitino and long vector multiplets. From table \ref{tab:KKVectorsSO8}, together with the analysis of \cite{Klebanov:2008vq}, one deduces that short and long gravitino and vector multiplets occurring at KK level $n$ with $\textrm{SU}(3) \times \textrm{U}(1)$ charges $[p,q]_{y_0}$ must have scaling dimensions:
\begin{eqnarray} \label{spectrumD=11GINO}
 \textrm{Short and long gravitino mult.} & : & E_0 = \tfrac12 + \sqrt{ \tfrac72 + \tfrac12 n (n+6) - \tfrac43 {\cal C}_2( p,q) +\tfrac12 y_0^2 }  \; , \\[5pt]
\label{spectrumD=11VEC}
 \textrm{Short and long vector mult.} & : & E_0 = \tfrac12 + \sqrt{ \tfrac{17}{4} + \tfrac12 n (n+6) - \tfrac43  {\cal C}_2( p,q) +\tfrac12 y_0^2 }  \; , \hspace{20pt}
\end{eqnarray}
in agreement with \cite{Malek:2020yue}. Here,
\begin{equation} \label{eq:SU3CasimirEigenv}
{\cal C}_2( p,q) \equiv \tfrac13 p^2 + \tfrac13 q^2 + \tfrac13 pq + p + q \; , 
\end{equation}
is the eigenvalue of the SU(3) quadratic Casimir operator. The data in table \ref{tab:KKVectorsSO8} are enough to infer the results  (\ref{spectrumD=11GINO}), (\ref{spectrumD=11VEC}) for all KK level $n$. A further calculation of the $n=3$ KK vector masses using (\ref{eq:KKVecMassMat}) is in agreement with these. Further, when the $\textrm{SU}(3) \times \textrm{U}(1)$ quantum numbers are restricted accordingly, (\ref{spectrumD=11GINO}), (\ref{spectrumD=11VEC}) reproduce the scaling dimensions of the short gravitino and vector multiplets given in \cite{Klebanov:2008vq}.

To summarise, the complete KK spectrum of the CPW solution contains the long and short $\textrm{OSp}(4|2) \times \textrm{SU}(3)$ multiplets specified in \cite{Klebanov:2008vq}. The dimension of the short multiplets, including the hypermultiplets, was given in that reference. The spectrum of (long and short) graviton mutliplets was given in \cite{Klebanov:2009kp}, and the spectrum of (long and short) gravitino and vector multiplets \cite{Malek:2020yue} is reproduced by (\ref{spectrumD=11GINO}) and (\ref{spectrumD=11VEC}) above. In the short cases, these correctly reduce to \cite{Klebanov:2008vq}.


\section{Complete KK spectra of AdS$_4$ solutions of type IIA} \label{sec:SpectraIIA}


Similar steps lead one to obtain the complete KK spectrum about the $\cN=2$ and $\cN=3$ AdS$_4$ solutions of massive type IIA supergravity \cite{Guarino:2015jca,Pang:2015vna,DeLuca:2018buk} that uplift from ISO(7) supergravity \cite{Guarino:2015jca,Gallerati:2014xra}. For these solutions, the spectrum at KK level $k=0$ 
\cite{Guarino:2015qaa,Gallerati:2014xra} and the graviton spectrum at all levels \cite{Pang:2017omp,Pang:2015rwd} are completely known. Here, I will give the complete supermultiplet structure and the scaling dimensions at all KK levels.

\subsection{Putative SO(7) structure of the KK spectra} \label{sec:General}

The massive type IIA AdS$_4$ solutions that uplift from ISO(7) supergravity have topologically $S^6$ internal spaces, equipped with $G$-invariant metrics for certain  subgroups $G$ of $\textrm{SO}(7)$. The spectrum for each solution should thus branch from SO(7) representations down to $G$ representations. Similarly to the SO(8) gauging case \cite{Englert:1983rn}, these putative SO(7) representations arise by tensoring the $\cN=8$ supergravity multiplet, herewith identified at the linearised level with KK level $k=0$, with the symmetric traceless representation $[k,0,0]$ of SO(7). Higgsing must also be taken into account: each massive graviton in a given SO(7) representation must eat a vector and a (pseudo)scalar in the same representation, which thereby disappear from the physical spectrum; each massive gravitino must eat a spin $1/2$ field; and each massive vector must eat a (pseudo)scalar. 

This exercise was carried out for the vectors in equation (\ref{eq:28xSymTrSO7}). More generally, table \ref{tab:SO7irrepsKKSpectrum} summarises the result for all the fields. The scalars and pseudoscalars in the table are respectively denoted $0^+$ and $0^-$. The KK level $k=0$ represented on the left table is actually $\bm{7}$ scalars short of being the $D=4$ $\cN=8$ supergravity multiplet. These scalars disappear from the spectrum, as they are St\"uckelberg and are eaten by the $\bm{7}$ vectors shown in the $k=0$ table. The latter always become massive at any AdS$_4$ vacuum. In turn, the massless vectors that gauge the residual symmetry $G$ always branch from the $\bm{21}$ vectors present at $k=0$.

The KK spectra about all the AdS$_4$ solutions in this class can be argued to have the SO(7) structure specified in table \ref{tab:SO7irrepsKKSpectrum}, even if dyonic ISO(7) supergravity does not have an $\cN=8$ SO(7)-invariant solution. From a bulk perspective, this SO(7) structure is in agreement with the ExFT approach of section \ref{sec:KKVecMassMat}. From the boundary point of view, an argument similar to that put forward \cite{Klebanov:2008vq} for the CPW solution \cite{Corrado:2001nv} can be made. The AdS$_4$ solutions under consideration of massive type IIA supergravity are dual to superconformal infrared fixed points of maximally supersymmetric three-dimensional Yang-Mills \cite{Guarino:2016ynd,Guarino:2019snw}. Despite its lack of conformal symmetry, and thus lack of a dual AdS$_4$ solution, the latter does have SO(7) R-symmetry. This SO(7) symmetry is thus inherited by all the infrared fixed points, necessarily branched out into representations of their corresponding flavour groups $G$. See \cite{Guarino:2019snw} for the holographic interpretation of the $k=0$ SO(7)-covariant fields in table \ref{tab:SO7irrepsKKSpectrum}.

 \begin{table}[]


\resizebox{\textwidth}{!}{

\begin{tabular}{lllll}
\hline
 spin & & SO(7) irrep & & SO(7) Dynkin labels 
            \\ \hline
$2$ && $\bm{1}$ && $[0,0,0]$ \\[4pt]
 $\frac32$ && $\bm{8}$ && $[0,0,1]$ \\[4pt]
 $1$ && $\bm{21}+\bm{7}$ && $[0,1,0] + [1,0,0] $ \\[4pt]
 $\frac12$ && $\bm{48}+\bm{8}$ &&  $[1,0,1] + [0,0,1] $ \\[4pt]
 $0^+$ && $\bm{27}+\bm{1}$ && $[2,0,0]+[0,0,0]$ \\[4pt]
 $0^-$ && $\bm{35}$ && $[0,0,2]$   \\
 \hline
\end{tabular}

\qquad

\begin{tabular}{lll}
\hline
 spin  & & SO(7) Dynkin labels 
            \\ \hline
$2$ && $[k,0,0]$ \\[4pt]
 $\frac32$ &&  $[k,0,1]+[k-1,0,1]$ \\[4pt]
 $1$ &&  $[k,1,0] + [k-1,0,2]+ [k-2,1,0] +[k+1,0,0] + [k-1,1,0] + [k-1,0,0] $ \\[4pt]
 $\frac12$ &&   $[k+1,0,1] + [k-1,1,1]+ [k-2,1,1] +[k-2,0,1] + [k,0,1] + [k-1,0,1] $ \\[4pt]
 $0^+$ &&  $[k+2,0,0] + [k,0,0]+ [k-2,2,0] +[k-2,0,0] $ \\[4pt]
 $0^-$ &&  $[k,0,2] + [k-1,1,0]+ [k-2,0,2] $   \\
 \hline
\end{tabular}
}

\caption{\footnotesize{States in SO(7) representations at KK level $k=0$ (left) and $k=0 , 1, 2 , \ldots$ (right) that compose the KK towers for AdS$_4$ solutions of massive IIA that uplift from ISO(7) supergravity. Representations with negative Dynkin labels are absent. For a solution with residual symmetry $G \subset \textrm{SO}(7)$, the spectrum organises itself in the representation of $G$ that branch from these SO(7) representations.}\normalsize}
\label{tab:SO7irrepsKKSpectrum}
\end{table}

\newpage


\subsection{Spectrum of the $\cN=2$ solution } \label{sec:N=2SpectrumIIA}


As has been just discussed, the KK spectrum of AdS$_4$ solutions with symmetry $G \subset \textrm{SO}(7)$ comes in the representations of $G$ that arise by branching the SO(7) representations in table \ref{tab:SO7irrepsKKSpectrum}. If, in addition, the solution preserves $\cN$ supersymmetries, the spectrum must also arrange itself into OSp$(4|\cN)$ supermultiplets. 

Let us go through the details for the $\cN=2$ $\textrm{SU}(3) \times \textrm{U}(1)$-invariant solution \cite{Guarino:2015jca}. At fixed KK level $k$, the SO(7) representations in table \ref{tab:SO7irrepsKKSpectrum}  for fields of each spin must be branched out under  $\textrm{SU}(3) \times \textrm{U}(1) \subset \textrm{SO}(7)$. The U(1) factor in the residual symmetry group corresponds to the R-symmetry and thus belongs to OSp$(4|2)$. The spectrum must thus be arranged in $\textrm{OSp} (4|2) \times \textrm{SU}(3)$ representations: at fixed KK level, fields of different spin and U(1) R-charge but the same SU(3) Dynkin labels $[p,q]$ must be allocated into OSp$(4|2)$ supermultiplets. The tables in appendix A of \cite{Klebanov:2008vq} come in very handy to carry out this exercise. I follow their notation for the $\textrm{OSp} (4|2)$ supermultiplets, only with a subindex $2$ attached in order to emphasise that these are $\cN=2$. This allocation into $\textrm{OSp} (4|2)$ supermultiplets proceeds from higher maximum spins to lower, as follows.

Firstly, the spin $s=2$ states are assigned to short, SGRAV$_2$ (or massless, MGRAV$_2$, for $k=0$), or long, LGRAV$_2$, graviton multiplets. Then, fields of lower spins in the same SU(3) representations are used to complete these supermultiplets. Secondly, The $s=3/2$ fields that were left unassigned to SGRAV$_2$ or LGRAV$_2$ multiplets are then ascribed to short, SGINO$_2$ or long, LGINO$_2$, gravitino multiplets. Again, fields of lower spin with the same Dynkin labels $[p,q]$ are then used to fill out these multiplets. Thirdly, the $s=1$ fields that still remain unassigned to the previous supermultiplets are allocated into short, SVEC$_2$ (or massless, MVEC$_2$, for $k=0$), or long, LVEC$_2$, vector multiplets. Spin $1/2$ fermions and scalars are then used to complete these supermultiplets. Finally, the remaining  $s=1/2$ fermions and scalars are assigned to hypermultiplets, HYP$_2$. 

The resulting multiplet structure is the following. At KK level $k=0$, this analysis was already carried out in \cite{Pang:2017omp}, and here I simply import the results summarised in table 3 therein. There is one real MGRAV$_2$ and one real MVEC$_2$, which are respectively a singlet and an $\bm{8}_0$ of $\textrm{SU}(3) \times \textrm{U}(1)$. The former corresponds to the $\cN=2$ pure supergravity multiplet and the latter contains the massless gauge fields in the adjoint of the residual symmetry group SU(3). Also at KK level $k=0$ there is one SGINO$_2$ and one HYP$_2$, both of them complex, together with their complex conjugates. KK level $k=0$ is completed with a real SU(3)-singlet\footnote{Consistent truncations of massive type IIA supergravity down to some of these $D=4$ multiplets are known at the full non-linear level. A non-linear consistent truncation to the MGRAV$_2$ was shown to exist in \cite{Varela:2019vyd}, in agreement with the general arguments of \cite{Gauntlett:2007ma,Cassani:2019vcl}. A similar result holds \cite{Larios:2019lxq} for the CPW solution. When non-linear interactions are restored, the MGRAV$_2$ and MVEC$_2$ furnish the SU(3)--invariant sector \cite{Guarino:2015qaa} of $D=4$ $\cN=8$ ISO(7) supergravity. This was explicitly embedded in type IIA in \cite{Varela:2015uca}. Of course, the entire KK level $k=0$ arises upon non-linear consistent truncation \cite{Guarino:2015jca,Guarino:2015vca} on $S^6$, similarly to the $D=11$ on $S^7$ case \cite{deWit:1986iy}.} LVEC$_2$. At each KK level starting at $k \geq 1$, there exists one, and only one, complex short OSp$(4|2)$ multiplet, and its complex conjugate, of each possible type, with $\textrm{SU}(3) \times \textrm{U}(1)$ charges $[p,q]_{y_0}$ fixed in terms of $k$. The list of short multiplets present in the spectrum is shown in table \ref{tab:ShortN=2Spectrum}. For completeness, the table also shows for each multiplet its scaling dimension $E_0$. For short multiplets, this is fixed in terms of the R-charge $y_0$ (see {\it e.g.} appendix A of \cite{Klebanov:2008vq}). All other multiplets are long. For example, there is a couple of infinite towers of long graviton multiplets with $\textrm{SU}(3) \times \textrm{U}(1)$ quantum numbers $[p,q]_{y_0}$ given by
\begin{equation}
\textrm{LGRAV}_2 \, : \quad  \bigoplus_{\ell=0}^{k-1} \bigoplus_{p=0}^{\ell} \, [p,\ell-p]_{\frac23(\ell-2p)} \, \oplus \, 
\bigoplus_{p=1}^{k-1}  \,  [p,k-p]_{\frac23(k-2p)}  \; , 
\end{equation}
with the first tower present for all $k \geq 1$ and the second kicking in at $k \geq 2$. The OSp$(4|2) \times \textrm{SU}(3)$ structure of the KK spectrum up to level $k=3$ is summarised in tables \ref{tab:multipletsatlevel0}--\ref{tab:multipletsatlevel3} below.

\begin{table}[]
\begin{center}
\begin{tabular}{lclcl} 					 \hline
SGRAV$_2 $ 	& : &  	$[k,0]_{-\frac{2k}{3}} + [0,k]_{+\frac{2k}{3}}$ \;   , &  & $E_0 = \tfrac{2k}{3} + 2$ 		\\[5pt] 
SGINO$_2 $ 	& : &  	$[k,1]_{-\frac{2k+1}{3}}  + [1,k]_{+\frac{2k+1}{3}} $ \; ,  &   & $E_0 =\tfrac{2k}{3} + \tfrac{11}{6} $ 		\\[5pt]
SVEC$_2 $ 	& :  &  	$[k+1,1]_{-\frac{2k}{3}}  + [1,k+1]_{+\frac{2k}{3}} $ \; ,   &  & $E_0 =\tfrac{2k}{3} + 1$ 		\\[5pt]
HYP$_2 $ 	& :  &  	$[k+2,0]_{-\frac{2}{3}(k+2)}  + [0,k+2]_{+\frac{2}{3}(k+2)} $ \; ,   &  & $E_0 =\tfrac{2k}{3} + \tfrac43$ 		\\ \hline
\end{tabular}
\caption{\footnotesize{The OSp$(4|2)$ short spectrum at KK level $k=0 , 1 , \ldots $ for the $\cN=2$ type IIA solution, in $[p,q]_{y_0}$ representations of $\textrm{SU}(3) \times \textrm{U}(1)$. For $k=0$, there is only one  $[0,0]_0$ graviton and only one $[1,1]_0$ vector multiplets, both of them massless.}\normalsize}
\label{tab:ShortN=2Spectrum}
\end{center}
\end{table}

With the spectrum allocated into $\textrm{OSp}(4|2) \times \textrm{SU}(3)$ representations, the conformal dimension $E_0$ for each multiplet remains to be given for the long multiplets. Unlike for the short ones, $E_0$ is not fixed by group theory in terms of the R-charge $y_0$, and is only required to obey a unitarity bound.  Thus, $E_0$ needs to be computed independently in these cases. By an argument similar to that made in section \ref{sec:N=2Spectrum11D} for CPW, the knowledge of the graviton spectrum \cite{Pang:2017omp}, together with the above group theory analysis and the individual KK vector spectra computed in section \ref{sec:KKVecMassMat}, is enough to reconstruct the entire, complete KK spectrum and all possible values of $E_0$.

The KK vector spectrum for the $\cN=2$ solution, computed from the mass matrix (\ref{eq:KKVecMassMat}), is presented for the first three KK levels, $k=0,1,2$, in the entry labelled as $\cN=2$ U(3) in table \ref{tab:KKVectorsISO7} of appendix \ref{sec:SpecificSpectra}. Firstly, the multiplicities shown in the table are compatible with the $\textrm{OSp}(4|2) \times \textrm{SU}(3)$  group theory just described. Secondly, all individual vectors that enter short multiplets do indeed have their masses fixed by the conformal dimension of those multiplets as follows from table \ref{tab:ShortN=2Spectrum}. Thirdly, there exist vector masses in the tables compatible with those predicted by the KK graviton analysis of \cite{Pang:2017omp}: these are the spin-1 states that accompany the spin-2 states into long and short graviton multiplets. The remaining individual vectors must arrange themselves in long gravitino and long vector multiplets. From the data in the table, one deduces that short or long gravitino and vector multiplets occurring at KK level $k$ with $\textrm{SU}(3) \times \textrm{U}(1)$ charges $[p,q]_{y_0}$ must have scaling dimensions:
\begin{eqnarray} \label{eq:SLGRAV2}
 \textrm{Short and long graviton mult.} & : & E_0 = \tfrac12 + \sqrt{ \tfrac94 + \tfrac23 k (k+5) - \tfrac43 \, {\cal C}_2( p,q) +\tfrac12 y_0^2 } \; ,   \hspace{10pt} \\[5pt]
\label{eq:SLGINO2}
 \textrm{Short and long gravitino mult.}  & : & E_0 = \tfrac12 + \sqrt{ \tfrac72 + \tfrac23 k (k+5) - \tfrac43 \, {\cal C}_2( p,q) +\tfrac12 y_0^2 } \; ,     \hspace{10pt}  \\[5pt]
\label{eq:SLVEC2}
 \textrm{Short and long vector mult.} & : & E_0 = \tfrac12 + \sqrt{ \tfrac{17}{4} + \tfrac23 k (k+5) - \tfrac{4}{3} \, {\cal C}_2( p,q) +\tfrac12 y_0^2 }    \; ,   \hspace{10pt}
\end{eqnarray}
with ${\cal C}_2( p,q)$ given in (\ref{eq:SU3CasimirEigenv}). The dimension $E_0$ in (\ref{eq:SLGRAV2}) for graviton multiplets has been imported from (3.1) of \cite{Pang:2017omp} with $n_{\textrm{there}} = k_{\textrm{here}}$, $\ell_{\textrm{there}} = p_{\textrm{here}}+ q_{\textrm{here}}$ and $y_{0\textrm{here}} = \frac23(  \ell_{\textrm{there}} - 2 p_{\textrm{there}} )$. The dimensions  (\ref{eq:SLGINO2}), (\ref{eq:SLVEC2}) for gravitino and vector multiplets have been deduced from table \ref{tab:KKVectorsISO7} and successfully cross-checked at level $k=3$. Further, when the $\textrm{SU}(3) \times \textrm{U}(1)$ quantum numbers $[p,q]_{y_0}$ are restricted in terms of $k$ as in table \ref{tab:ShortN=2Spectrum}, the dimensions (\ref{eq:SLGRAV2})--(\ref{eq:SLVEC2}) correctly reproduce the short multiplet dimensions given in that table. At this stage, the dimensions of all short and long multiplets that occur in the spectrum have been identified.

Tables \ref{tab:multipletsatlevel0}--\ref{tab:multipletsatlevel3} below show the OSp$(4|2) \times \textrm{SU}(3)$ representations present in the spectrum at KK levels $k=0,1,2$. At fixed KK level, OSp$(4|2)$ supermultiplets with the same SU(3) Dynkin labels $[p,q]$ are grouped up in the same cell. The scaling dimension $E_0$ of each supermultiplet, computed via (\ref{eq:SLGRAV2})--(\ref{eq:SLVEC2}) and table \ref{tab:ShortN=2Spectrum}, and its R-charge $y_0$, computed with group theory as described above, are shown as a label $(E_0)_{y_0}$ next to each entry. A label $2 \times (E_0)_{y_0}$ indicates that there are two such multiplets. States in SU(3) representations with $[p,q]$ quantum numbers such that $q > p$ appear in the tables as ``conjugate to $[q,p]$" representations; these have their R-charges negated. For example, in table \ref{tab:multipletsatlevel0} at $k=0$ there is a $[0,1]$ SGINO$_2$ with $(E_0)_{y_0} = \left( \tfrac{11}{6} \right)_{-\frac{1}{3}}$. The format of these tables has been kindly borrowed from \cite{Klebanov:2008vq}.

\vspace{30pt}


\begin{table}[H]
\begin{center}
{\footnotesize
\begin{tabular}{|p{30mm}|p{30mm}|p{30mm}|} 					\hline
$[0,0]$ 				& 	$[0,1]$ 				& 	$[0,2]$ 			\\[1pt]
MGRAV$_2$ $\left( 2 \right)_0$ 			& conj.~to $[1,0]$ 	& conj.~to $[2,0]$ 	\\
LVEC$_2$  $\left( \tfrac12 +\tfrac{\sqrt{17}}{2} \right)_0$ 		& & 					\\[5pt] \hline
$[1,0]$ 				& 	$[1,1]$ 			\\[1pt]
SGINO$_2$ $\left( \tfrac{11}{6} \right)_{+\frac{1}{3}} $ & MVEC$_2$ $\left( 1  \right)_0$ 			\\[5pt] \cline{1-2}
$[2,0]$ 			\\[1pt]
HYP$_2$ $\left( \tfrac{4}{3} \right)_{-\frac{4}{3}}$ 	\\[5pt] \cline{1-1}
\end{tabular}
}
\caption{\footnotesize{$\cN=2$ supermultiplets at KK level $k=0$.}\normalsize}
\label{tab:multipletsatlevel0}
\end{center}
\end{table}

\vspace{60pt}


\begin{table}[H]
\begin{center}
{\footnotesize
\begin{tabular}{|p{35mm}|p{35mm}|p{35mm}|p{35mm}|} \hline
$[0,0]$ 							& 	$[0,1]$ 				& 	$[0,2]$ 			& 	$[0,3]$ \\[1pt]
LGRAV$_2$ $\left( 3 \right)_0$			& conj.~to $[1,0]$ 	& conj.~to $[2,0]$ 	& conj.~to $[3,0]$ \\
LVEC$_2$ $\left( \tfrac12+\tfrac{\sqrt{33}}{2} \right)_0$	&        	&      	&  \\[5pt] \hline
$[1,0]$ 					& 	$[1,1]$ 					& 	$[1,2]$ 			\\[1pt]
SGRAV$_2$ $\left( \tfrac83 \right)_{-\frac23}$ 	& SGINO$_2$ $\left( \tfrac52 \right)_{\pm1}$	& conj.~to $[2,1]$ 		\\
LGINO$_2$ $\left( \tfrac12+\tfrac{2\sqrt{13}}{3} \right)_{+\frac13}$		& LVEC$_2$ $\left( \tfrac12+\tfrac{\sqrt{17}}{2} \right)_{0}$ 		&                    			\\
LVEC$_2$ $\left( \tfrac12+\tfrac{\sqrt{241}}{6} \right)_{-\frac23}$			&           						&                    			\\[5pt]  \cline{1-3}
$[2,0]$ 			& 	$[2,1]$ 				\\[1pt]
LGINO$_2$ $\left( \tfrac12+\tfrac{2\sqrt{7}}{3} \right)_{-\frac13}$ 	& SVEC$_2$ $\left( \tfrac53 \right)_{-\frac23}$	 		\\[5pt] \cline{1-2}
$[3,0]$ \\[1pt]
HYP$_2$ $\left( 2 \right)_{-2}$ \\[5pt] \cline{1-1}
\end{tabular}
}
\caption{\footnotesize{$\cN=2$  supermultiplets at KK level $k=1$.}}
\label{tab:multipletsatlevel1}
\end{center}
\end{table}

\vspace{40pt}


\begin{table}[H]
\begin{center}
{\footnotesize
\begin{tabular}{|p{35mm}|p{35mm}|p{30mm}|p{20mm}|p{20mm}|} \hline
$[0,0]$ 								& 	$[0,1]$ 			& 	$[0,2]$ 			& 	$[0,3]$			& 	$[0,4]$ 		\\[1pt]
LGRAV$_2$ $\left( \tfrac12+ \tfrac{\sqrt{417}}{6} \right)_0$								& 	conj. to [1,0] 			 	& 	conj. to [2,0]				&  conj. to [3,0]					&		conj. to [4,0]		\\[5pt]
LVEC$_2$ $2 \times \left( \tfrac12+ \tfrac{\sqrt{489}}{6} \right)_0$				& 		 	& 			& 			&		\\[5pt] \hline
$[1,0]$ 										& 	$[1,1]$ 						& 	$[1,2]$ 			&	$[1,3]$			\\[1pt]
LGRAV$_2$ $\left( \tfrac{11}{3} \right)_{-\frac23}$				& 	LGRAV$_2$ $\left( \tfrac12 + \tfrac{\sqrt{273}}{6} \right)_{0}$					& 					&					\\[4pt]
LGINO$_2$ $2 \times \left( \tfrac{23}{6} \right)_{\frac13}$		& 	LGINO$_2$ $\left( \tfrac12 + \tfrac{2\sqrt{21}}{3} \right)_{\pm 1 }$	&      conj. to [2,1]		&	conj. to [3,1]		\\[5pt]
LVEC$_2$ $\left( \tfrac12 + \tfrac{\sqrt{433}}{6} \right)_{-\frac23}$ 								& 	LVEC$_2$ $ 2 \times \left( \tfrac12+ \tfrac{\sqrt{345}}{6} \right)_0$		&      					&					\\[8pt] \cline{1-4}
$[2,0]$ 					&	$[2,1]$ 						&	$[2,2]$		\\[1pt]
SGRAV$_2$ $\left( \tfrac{10}{3} \right)_{-\frac43}$		&	SGINO$_2$  $\left( \tfrac{19}{6} \right) _{-\frac53}$ 	&	LVEC$_2$ $\left( \tfrac12 +\tfrac{ \sqrt{105}}{6} \right)_0$  \\
LGINO$_2$ $\left( \tfrac12 + \tfrac{2\sqrt{19}}{3} \right)_{-\frac13}$	 	&	LGINO$_2$ $\left( \tfrac12 + \tfrac{2\sqrt{13}}{3}  \right)_{+\frac13}$		&   \\
LVEC$_2$ $\left( \tfrac{11}{3} \right)_{-\frac43}$		 	&	LVEC$_2$ $\left( \tfrac12 + \tfrac{\sqrt{241}}{6}  \right)_{-\frac23}$	&   \\[4pt]
LVEC$_2$ $\left( \tfrac12 + \tfrac{\sqrt{337}}{6} \right)_{+\frac23}$	&						&   \\[6pt] \cline{1-3}
$[3,0]$ 				&	$[3,1]$			\\[1pt]
LGINO$_2$ $\left( \tfrac12 + \tfrac{4\sqrt{3}}{3} \right)_{-1}$	 		&	SVEC$_2$ $\left( \tfrac73 \right) _{-\frac43}$	\\[8pt]  \cline{1-2}
$[4,0]$ 			\\[1pt]
HYP$_2$ $\left( \tfrac{8}{3} \right)_{-\frac83}$	\\[5pt] \cline{1-1}
\end{tabular}
}
\caption{\footnotesize{$\cN=2$  supermultiplets at KK level $k=2$.}\normalsize}
\label{tab:multipletsatlevel2}
\end{center}
\end{table}


\begin{table}[H]
\begin{center}
\begin{sideways}
{\scriptsize
\begin{tabular}{|l|l|l|l|l|l|} \hline
$[0,0]$ & $[0,1]$ & $[0,2]$ & $[0,3]$ & $[0,4]$ & $[0,5]$ \\
LGRAV$_2$ $\left( \tfrac{1}{2} +\tfrac{\sqrt{73}}{2} \right)_0$ & conj. to $[1,0]$ & conj. to $[2,0]$ & conj. to $[3,0]$ & conj. to $[4,0]$ & conj. to $[5,0]$ \\[1pt]
LVEC$_2$ $2 \times \left( 5 \right)_0$ & & & & &        \\[5pt] \hline
$[1,0]$ & $[1,1]$ & $[1,2]$ & $[1,3]$ & $[1,4]$ \\[1pt]
LGRAV$_2$ $\left(\tfrac12 +\tfrac{\sqrt{601}}{6} \right)_{-\frac{2}{3}}$   & LGRAV$_2$ $\left( \tfrac12 + \tfrac{\sqrt{57}}{2} \right)_0$ & conj. to $[2,1]$ & conj. to $[3,1]$ & conj. to $[4,1]$ \\[7pt]
LGINO$_2$ $2 \times \left( \tfrac12 + \tfrac{4\sqrt{10}}{3} \right)_{+\frac{1}{3}}$  & LGINO$_2$ $2 \times \left( \tfrac{9}{2} \right)_{\pm 1}$ & & & \\
LVEC$_2$ $2 \times \left( \tfrac12 + \tfrac{\sqrt{673}}{6} \right)_{-\frac{2}{3}}$ & LVEC$_2$ $2 \times \left( \tfrac12 + \tfrac{\sqrt{65}}{2} \right)_{0}$  & & & \\[8pt] \cline{1-5}
$[2,0]$ & $[2,1]$ & $[2,2]$ & $[2,3]$ \\[1pt]
LGRAV$_2$ $\left( \tfrac{13}{3} \right)_{-\frac{4}{3}}$  & LGRAV$_2$ $\left( \tfrac12 + \tfrac{\sqrt{409}}{6} \right)_{-\frac{2}{3}}$  & LGINO$_2$  $\left( \tfrac12 + \tfrac{2\sqrt{21}}{3} \right)_{\pm 1}$  & conj. to $[3,2]$ \\[8pt]
LGINO$_2$ $2 \times \left( \tfrac12 + \tfrac{2\sqrt{34}}{3} \right)_{-\frac{1}{3}}$  & LGINO$_2$ $\left( \tfrac12 + \tfrac{4\sqrt{7}}{3} \right)_{+\frac{1}{3}}$  ,  $\left( \tfrac12 + \tfrac{2\sqrt{31}}{3} \right)_{-\frac{5}{3}}$  & LVEC$_2$ $\left( \tfrac12 + \tfrac{\sqrt{345}}{6} \right)_0$ & \\[10pt]
LVEC$_2$  $\left( \tfrac12 + \tfrac{\sqrt{601}}{6} \right)_{-\frac{4}{3}}$ , $\left( \tfrac12 + \tfrac{\sqrt{577}}{6} \right)_{+\frac{2}{3}}$    & LVEC$_2$ $2 \times \left( \tfrac12 + \tfrac{\sqrt{481}}{6} \right)_{-\frac{2}{3}}$   , $\left( \tfrac12 + \tfrac{\sqrt{505}}{6} \right)_{+\frac{4}{3}}$   & & \\[8pt]\cline{1-4}
$[3,0]$ & $[3,1]$ & $[3,2]$ \\[1pt]
SGRAV$_2$ $\left( 4 \right)_{-2}$ & SGINO$_2$ $\left( \tfrac{23}{6} \right)_{-\frac{7}{3}}$ & LVEC$_2$ $\left( \tfrac12 + \tfrac{\sqrt{193}}{6} \right)_{-\frac{2}{3}}$ \\[4pt]
LGINO$_2$ $\left( \tfrac12 +2\sqrt{3} \right)_{-1}$ & LGINO$_2$ $\left( \tfrac12 + \tfrac{2\sqrt{19}}{3} \right)_{-\frac{1}{3}}$ & \\[4pt]
LVEC$_2$ $\left( \tfrac12 + \tfrac{\sqrt{57}}{2} \right)_{-2}$ , $\left( 4 \right)_{0}$   &  LVEC$_2$ $\left( \tfrac{11}{3} \right)_{-\frac{4}{3}}$  & \\[8pt] \cline{1-3}
$[4,0]$ & $[4,1]$ \\[1pt]
LGINO$_2$ $\left( \tfrac12 + \tfrac{2\sqrt{19}}{3} \right)_{-\frac{5}{3}}$  & SVEC$_2$ $\left( 3 \right)_{-2}$ \\[5pt] \cline{1-2}
$[5,0]$ \\[1pt]
HYP$_2$ $\left( \tfrac{10}{3} \right)_{-\frac{10}{3}}$  \\[8pt] \cline{1-1}
\end{tabular}
}
\end{sideways}
\caption{\footnotesize{$\cN=2$ supermultiplets at KK level $k=3$.}\normalsize}
\label{tab:multipletsatlevel3}
\end{center}

\end{table}

\newpage

 \begin{table}

\centering

\scriptsize{

\begin{tabular}{llccclcl}
\hline
\multicolumn{2}{c}{$\Delta$}  && $k $  &&   $\textrm{SU}(3) \times \textrm{U}(1) $  && $\cN=2$ supermultiplet
            \\ \hline
$1$ & $1.00$ & &  0   &&  $[1,1]_0  $   && MVEC$_2$ $\left( 1 \right)_0  $ \\[5pt]
%
 %
 %
 %
 $\tfrac43$ & $1.33$  &&  0   &&  $[2,0]_{-\tfrac43}  + \textrm{c.c.} $   && HYP$_2$ $\left( \tfrac43 \right)_{-\tfrac43}  + \textrm{c.c.} $ \\[5pt] 
%
 %
 %
 $\tfrac53$ & $1.67$  &&  1   &&  $[2,1]_{-\tfrac23}  + \textrm{c.c.} $   && SVEC$_2$ $\left( \tfrac53 \right)_{-\tfrac23} + \textrm{c.c.}  $ \\[5pt] 
%
 %
 %
$2$ & $2.00$ & &  0   &&  $[1,1]_0  $   && MVEC$_2$ $\left( 1 \right)_0  $ \\[5pt] 
 %
%
 %
 %
 $2$ & $2.00$  &&  1   &&  $[3,0]_{-2}  + \textrm{c.c.} $   && HYP$_2$ $\left( 2 \right)_{-2} + \textrm{c.c.}  $ \\[5pt] 
%
 %
 %
$ \tfrac12 + \tfrac{\sqrt{105}}{6} $ & $2.21$ & &  2   &&  $[2,2]_{0}   $   && LVEC$_2$ $\left( \tfrac12 + \tfrac{\sqrt{105}}{6} \right)_{0}  $ \\[5pt] 
 %
%
 %
 %
$\tfrac73$ & $2.33$ & &  0   &&  $[1,0]_{-\frac23}  + \textrm{c.c.} $   && SGINO$_2$ $\left( \tfrac{11}{6} \right)_{\frac13} + \textrm{c.c.} $ \\[5pt] 
 %
 %
 %
 $\tfrac73$ & $2.33$  &&  0   &&  $[2,0]_{\tfrac23}  + \textrm{c.c.} $   && HYP$_2$ $\left( \tfrac43 \right)_{-\tfrac43} + \textrm{c.c.}  $ \\[5pt] 
%
 %
 %
$\tfrac73$ & $2.33$ & &  2   &&  $[3,1]_{-\frac43}  + \textrm{c.c.} $   && SVEC$_2$ $\left( \tfrac{7}{3} \right)_{-\frac43} + \textrm{c.c.} $ \\[5pt] 
 %
  %
 %
 %
$ \tfrac12 + \tfrac{\sqrt{17}}{2} $ & $2.56$ & &  0   &&  $[0,0]_{0}   $   && LVEC$_2$ $\left( \tfrac12 + \tfrac{\sqrt{17}}{2} \right)_{0}  $ \\[5pt] 
 %
%
 %
 %
$ \tfrac12 + \tfrac{\sqrt{17}}{2} $ & $2.56$ & & 1   &&  $[1,1]_{0}   $   && LVEC$_2$ $\left( \tfrac12 + \tfrac{\sqrt{17}}{2} \right)_{0}  $ \\[5pt] 
 %
 %
 %
 $\tfrac83$ & $2.67$  &&  1   &&  $[2,1]_{-\tfrac23}  + \textrm{c.c.} $   && SVEC$_2$ $\left( \tfrac53 \right)_{-\tfrac23} + \textrm{c.c.}  $ \\[5pt] 
%
 %
 %
 $\tfrac83$ & $2.67$  &&  1   &&  $[2,1]_{+\tfrac43}  + \textrm{c.c.} $   && SVEC$_2$ $\left( \tfrac53 \right)_{-\tfrac23} + \textrm{c.c.}  $ \\[5pt] 
%
%
%
 %
 %
 $\tfrac83$ & $2.67$  &&  2   &&  $[4,0]_{-\tfrac83}  + \textrm{c.c.} $   && HYP$_2$ $\left( \tfrac83 \right)_{-\tfrac83} + \textrm{c.c.}  $ \\[5pt] 
%
 %
 %
$ 1 + \tfrac{2\sqrt{7}}{3} $ & $2.76$ & & 1   &&  $[2,0]_{-\frac43} + \textrm{c.c.}  $   && LGINO$_2$ $\left( \tfrac12 + \tfrac{2\sqrt{7}}{3} \right)_{-\tfrac13}  $ \\[5pt] 
 %
%
%
 %
 %
$ 1 + \tfrac{2\sqrt{7}}{3} $ & $2.76$ & & 1   &&  $[2,0]_{+\frac23} + \textrm{c.c.}  $   && LGINO$_2$ $\left( \tfrac12 + \tfrac{2\sqrt{7}}{3} \right)_{-\tfrac13}  $ \\[5pt] 
 %
 %
 %
 %
$ \tfrac12 + \tfrac{\sqrt{193}}{6} $ & $2.81$ & &  3   &&  $[3,2]_{-\tfrac23}   $   && LVEC$_2$ $\left( \tfrac12 + \tfrac{\sqrt{193}}{6} \right)_{-\tfrac23}  $ \\[5pt] 
 %
%
 %
 %
$3$ & $3.00$ & &  1   &&  $[1,1]_{0}   $   && SGINO$_2$ $\left( \tfrac{5}{2} \right)_{-1}  $ \\[5pt] 
 %
 %
%
 %
 %
$3$ & $3.00$ & &  1   &&  $[1,1]_{0}   $   && SGINO$_2$ $\left( \tfrac{5}{2} \right)_{+1}  $ \\[5pt] 
 $3$ & $3.00$  &&  1   &&  $[3,0]_{0}  + \textrm{c.c.} $   && HYP$_2$ $\left( 2 \right)_{-2} + \textrm{c.c.}  $ \\[5pt] 
%
%
 %
 %
$3$ & $3.00$ & &  3   &&  $[4,1]_{-2}   $   && SVEC$_2$ $\left( 3 \right)_{-2}  $ \\[5pt] 
 \hline
\end{tabular}

}\normalsize

\caption{\scriptsize{All KK scalars with dimension $\Delta \leq 3$ around the $\cN=2$ AdS$_4$ type IIA solution.  }\normalsize}
\label{tab:N=2ISO7Scalars}
\end{table}

The spectrum of KK scalars can be deduced from the above results. Table \ref{tab:N=2ISO7Scalars} lists all the scalars with scaling  dimensions $\Delta$ less than or equal to 3. The table includes the analytical value of $\Delta$ together with a convenient numerical approximation. Also shown in the table is the KK level $k$ at which each scalar appears, as well as its $\textrm{SU}(3) \times \textrm{U}(1)$ charges $[p,q]_r$. The OSp$(4|2)$ supermultiplet with dimension and R-charge $(E_0)_{y_0}$, at the same KK level $k$ and with the same SU(3) charges $[p,q]$, that contains each scalar is also shown. The tables in appendix A of \cite{Klebanov:2008vq} are useful to derive $\Delta$ and $r$ from $E_0$ and $y_0$. These will only match if the scalar is the superconformal primary of the multiplet. The scalars in table \ref{tab:N=2ISO7Scalars} are dual to relevant ($\Delta < 3$) or classically marginal ($\Delta = 3$) operators in the dual field theory. All scalars with $\Delta \leq 3$ turn out to arise at KK levels $k=0,1,2,3$. Each of these KK levels contain scalars dual to irrelevant ($\Delta >3$) operators as well. At KK levels $k \geq 4$, all scalars are dual to irrelevant operators.

\newpage 


\subsection{Spectrum of the $\cN=3$ solution} \label{sec:N=3SpectrumIIA}


The KK spectrum of the $\cN=3$ AdS$_4$ solution of \cite{Pang:2015vna,DeLuca:2018buk} can be obtained similarly. The residual symmetry group of this solution is the
$\textrm{SO}(4) \equiv \textrm{SO}(3)_\cR \times \textrm{SO}(3)_\cF$ 
subgroup of SO(7) defined through the embedding chain \cite{Gallerati:2014xra}
\begin{equation} \label{eq:embeddingSO3dSOrR}
\textrm{SO}(7) \, \supset \, 
  \textrm{SO}(3)^\prime \times \textrm{SO}(4)^\prime \equiv  \textrm{SO}(3)^\prime \times \textrm{SO}(3)_{\textrm{left}} \times \textrm{SO}(3)_{\textrm{right}} 
 \supset
  \textrm{SO}(3)_{\cR} \times \textrm{SO}(3)_{\cF} \; .
\end{equation}
Here, $\textrm{SO}(3)_{\cF} \equiv \textrm{SO}(3)_{\textrm{right}} $ is the right-handed component of $\textrm{SO}(4)^\prime$, and $\textrm{SO}(3)_{\cR}$ is the diagonal subgroup of $\textrm{SO}(3)^\prime \times \textrm{SO}(3)_{\textrm{left}}$. The group $\textrm{SO}(3)_{\cR}$ is the R-symmetry and is thus contained within OSp$(4|3)$, while $\textrm{SO}(3)_{\cF}$ is the flavour group of the dual field theory. The KK spectrum about this solution thus comes in $\textrm{OSp}(4|3) \times \textrm{SO}(3)_\cF$ representations. The representations of $\textrm{SO}(3)_{\cR} \times \textrm{SO}(3)_{\cF}$ given below are labelled $(j,h)$, in terms of the spins $j =0 ,\tfrac12 , 1 , \tfrac32 , 2 , \ldots$ and $h=0 ,\tfrac12 , 1 , \tfrac32 , 2 , \ldots$ of $\textrm{SO}(3)_{\cR}$  and $\textrm{SO}(3)_{\cF}$, such that the spin $j$ representation of $\textrm{SO}(3)_{\cR}$ is $(2j+1)$-dimensional, and similarly for  $\textrm{SO}(3)_{\cF}$. I follow \cite{Fre:1999gok} in referring to the R-symmetry $\textrm{SO}(3)_{\cR}$ spin $j$ as {\it isospin}.

At fixed KK level $k$, the SO(7) representations in table \ref{tab:SO7irrepsKKSpectrum}  for fields of each physical spin must be branched out under  (\ref{eq:embeddingSO3dSOrR}). Fields of different physical spin and SO$(3)_\cR$ isospin $j$ but the same SO$(3)_\cF$ flavour spin $h$ must be allocated into the same OSp$(4|3)$ supermultiplets. These supermultiplets have been summarised for convenience in appendix \ref{sec:N=3Supermultiplets} (with the isospin denoted therein as $j_0$). The allocation proceeds from higher maximum physical spins to lower spins, similar to the $\cN=2$ case discussed in detail in section \ref{sec:N=2SpectrumIIA}. In the present $\cN=3$ case, the process terminates by allocating the individual vectors that did not enter graviton or gravitino multiplets into vector multiplets, as there are no $\cN=3$ hypermultiplets. Further, the $\cN=3$ vector multiplets are necessarily short.

The resulting multiplet structure is the following. At KK level $k=0$ (given already in \cite{Gallerati:2014xra}), there is one MGRAV$_3$ and one MVEC$_3$, which are respectively a singlet and
a $(1,1)$ of $\textrm{SO}(3)_{\cR} \times \textrm{SO}(3)_{\cF}$. The former corresponds to the $\cN=3$ pure supergravity multiplet. This contains the massless graviton, the $\cN=3$ massless gravitini and the massless R-symmetry graviphotons which lie in the adjoint of $\textrm{SO}(3)_{\cR}$ and are singlets under $\textrm{SO}(3)_{\cF}$. The MVEC$_3$ multiplet contains the massless R-symmetry-singlet vectors, in the adjoint of $\textrm{SO}(3)_{\cF}$, that gauge the residual flavour group. There is also a $\left( \tfrac12 , \tfrac12 \right) $  SGINO$_3$ and a singlet\footnote{Nonlinear consistent truncations to theories containing some of these multiplets are known. The consistent truncation predicted by \cite{Gauntlett:2007ma,Cassani:2019vcl} to the pure $\cN=3$ gravity multiplet, MGRAV$_3$, was constructed in \cite{Varela:2019vyd}. A truncation that retains the singlet MGRAV$_3$ and the LGINO$_3$ was described in \cite{Guarino:2019jef,Guarino:2019snw}.} LGINO$_3$. At each KK level starting at $k \geq 1$, there exists one, and only one, short OSp$(4|3)$ multiplet of each possible type. The list of short multiplets present in the spectrum is given in table \ref{tab:ShortN=3Spectrum}. For completeness, the table also shows for each multiplet its scaling dimension $E_0$. This is fixed in terms of the isospin $j$ as reviewed in appendix \ref{sec:N=3Supermultiplets}. All other multiplets are long. The $\textrm{OSp}(4|3) \times \textrm{SO}(3)_\cF$ structure of the KK spectrum up to level $k=3$ is summarised in tables \ref{tab:N=3multipletsatlevel0}--\ref{tab:N=3multipletsatlevel3} below. All multiplets are real.

\begin{table}[]
\begin{center}
\begin{tabular}{lclcl} 					 \hline
SGRAV$_3 $ 	& : &  	$\left( k, 0 \right) $ \;   , &  & $E_0 = k+\tfrac{3}{2} $ 		\\[5pt]
SGINO$_3 $ 	& : &  	$ \left( \tfrac{k+1}{2} , \tfrac{k+1}{2} \right)  $ \; ,  &   & $E_0 =\tfrac{k+3 }{2}  $ 		\\[5pt]
SVEC$_3 $ 	& :  &  	$\left( \tfrac{k+2}{2} , \tfrac{k+2}{2} \right) $ \; ,   &  & $E_0 = \tfrac{k+2}{2}  $ 		\\
\hline
\end{tabular}
\caption{\footnotesize{The OSp$(4|3)$ short spectrum at KK level $k=0 , 1 , \ldots $ for the $\cN=3$ type IIA solution, in $(j, h)$ representations of $\textrm{SO}(3)_\cR \times \textrm{SO}(3)_\cF$. For $k=0$, the short graviton and vector multiplets become massless.}\normalsize}
\label{tab:ShortN=3Spectrum}
\end{center}
\end{table}

Now that the spectrum has been arranged into $\textrm{OSp}(4|3) \times \textrm{SO}(3)_\cF$ representations, the conformal dimension $E_0$ of the long multiplets remains to be given. Like in the previous cases, 
the knowledge of the graviton spectrum \cite{Pang:2015rwd}, together with the above group theory and the individual KK vector spectra computed from the mass matrix (\ref{eq:KKVecMassMat}), is enough to reconstruct the entire, complete KK spectrum and all possible values of $E_0$. The KK vector spectrum for the $\cN=3$ solution is presented for the first three KK levels, $k=0,1,2$, in the entry labelled as $\cN=3$ SO(4) in table \ref{tab:KKVectorsISO7} of appendix \ref{sec:SpecificSpectra}. Firstly, the multiplicities shown in the table are compatible with the $\textrm{OSp}(4|3) \times \textrm{SO}(3)_\cF$  group theory just described. Secondly, all individual vectors that enter short multiplets do indeed have their masses fixed by the conformal dimension of those multiplets as follows from table \ref{tab:ShortN=3Spectrum}. Thirdly, there exist vector masses in the tables compatible with those predicted by the KK graviton analysis of \cite{Pang:2015rwd}. The remaining individual vectors must arrange themselves in long gravitino multiplets. From the data in the table, one deduces that short and long graviton and gravitino multiplets occurring at KK level $k$ with $\textrm{SO}(3)_\cR \times \textrm{SO}(3)_\cF$ charges $(j,h)$ must have scaling dimensions:
{\setlength\arraycolsep{1pt}
\begin{eqnarray} \label{eq:SLGRAV3}
 \textrm{Short and long graviton mult.} & : & E_0 = \sqrt{ \tfrac94 + \tfrac12 k (k+5) + \tfrac12 \, j (j+1) -\tfrac32 \, h (h+1) }  ,   \hspace{20pt} \\[5pt]
\label{eq:SLGINO3}
 \textrm{Short and long gravitino mult.}  & : & E_0 =  \sqrt{ 3 + \tfrac12 k (k+5) + \tfrac12 \, j (j+1) -\tfrac32 \, h (h+1) }  .   \hspace{20pt}  
\end{eqnarray}
}The dimension $E_0$ in (\ref{eq:SLGRAV3}) corresponds, up to straightforward notational changes, to the expression given in appendix B of \cite{Pang:2017omp} for the graviton multiplet dimensions \cite{Pang:2015rwd}. The dimension (\ref{eq:SLGINO3}) for short and long gravitino multiplets has been deduced from table \ref{tab:KKVectorsISO7} and successfully cross-checked at level $k=3$. Further, when the $\textrm{SO}(3)_{\cR} \times \textrm{SO}(3)_{\cF}$ quantum numbers $(j,h)$ are restricted in terms of the KK level $k$ as in table \ref{tab:ShortN=3Spectrum}, the dimensions (\ref{eq:SLGRAV3}), (\ref{eq:SLGINO3}) correctly reproduce the short multiplet dimensions given in that table. 

Tables \ref{tab:N=3multipletsatlevel0}--\ref{tab:N=3multipletsatlevel3} below summarise the $\textrm{OSp}(4|3) \times \textrm{SO}(3)_\cF$ representations present in the spectrum up to KK level $k=3$. For each $\textrm{OSp}(4|3)$ supermultiplet at fixed $\textrm{SO}(3)_\cF$ quantum number $h$, its conformal dimension and SO$(3)_\cR$ isospin are indicated with a label $(E_0)_{j}$ next to each entry. Labels of the form $2 \times  (E_0)_{j}$ indicate that there are two such multiplets.


\newpage

\begin{landscape}


\begin{table}[H]
\begin{center}
{\tiny
\begin{tabular}{|c|c|c|}
\hline
 $h=0$ &  $h= \tfrac12$  & $h=1$
            \\ \hline
MGRAV$_3$ $\left( \tfrac32 \right)_0$ & SGINO$_3$ $\left( \tfrac32 \right)_{\frac{1}{2}}$ & MVEC$_3$ $\left( 1 \right)_1$ \\[4pt]
LGINO$_3$ $\left( \sqrt{3} \right)_{0}$ &  &  \\[1pt]
 \hline
\end{tabular}
}\normalsize
\caption{\footnotesize{$\cN=3$ supermultiplets at KK level $k=0$} \normalsize}
\label{tab:N=3multipletsatlevel0}
\end{center}
\end{table}

\vspace{10pt}


\begin{table}[H]
\begin{center}
{\tiny
\begin{tabular}{|c|c|c|c|}
\hline
 $h=0$ &  $h= \tfrac12$  & $h=1$ & $h=\tfrac32$
            \\ \hline
SGRAV$_3$ $\left( \tfrac52 \right)_1$ & LGRAV$_3$ $\left( \tfrac{3\sqrt{2}}{2} \right)_{\frac{1}{2}}$ & SGINO$_3$ $\left( 2 \right)_1$ & SVEC$_3$ $\left( \tfrac32 \right)_{\frac{3}{2}}$ \\[4pt]
LGINO$_3$ $\left( \sqrt{7} \right)_1$ & LGINO$_3$ $\left( \tfrac{\sqrt{21}}{2} \right)_{\frac{1}{2}}$ & LGINO$_3$ $\left( \sqrt{3} \right)_0$ &  \\[6pt]
 \hline
\end{tabular}
}\normalsize
\caption{\footnotesize{$\cN=3$ supermultiplets at KK level $k=1$} }\normalsize
\label{tab:N=3multipletsatlevel1}
\end{center}
\end{table}

\vspace{10pt}


\begin{table}[H]
\begin{center}
{\tiny
\begin{tabular}{|c|c|c|c|c|}
\hline
 $h=0$ &  $h= \tfrac12$  & $h=1$ & $h=\tfrac32$ & $h=2$
            \\ \hline
SGRAV$_3$ $\left( \tfrac72 \right)_2$ & LGRAV$_3$ $\left( \tfrac{\sqrt{34}}{2} \right)_{\frac{1}{2}}$ , $\left( \sqrt{10} \right)_{\frac{3}{2}}$  & LGRAV$_3$ $\left( \tfrac{\sqrt{29}}{2} \right)_1$ & SGINO$_3$ $\left( \tfrac52 \right)_{\frac{3}{2}}$  & SVEC$_3$ $\left( 2 \right)_2$  \\[6pt]
LGRAV$_3$ $\left( \tfrac{\sqrt{37}}{2} \right)_0$ & LGINO$_3$ $\left( \tfrac{\sqrt{37}}{2} \right)_{\frac{1}{2}}$ , $\left( \tfrac{\sqrt{43}}{2} \right)_{\frac{3}{2}}$ & LGINO$_3$ $2 \times  \left(  2 \sqrt{2} \right)_1$ & LGINO$_3$ $\left( \tfrac{\sqrt{19}}{2} \right)_{\frac{1}{2}}$  &   \\[6pt]
LGINO$_3$ $2 \times \left(  \sqrt{10} \right)_0$ , $\left(  \sqrt{13} \right)_2$   &   &   &  &  \\[3pt]
 \hline
\end{tabular}
}\normalsize
\caption{\footnotesize{$\cN=3$ supermultiplets at KK level $k=2$}\normalsize}
\label{tab:N=3multipletsatlevel2}
\end{center}
\end{table}

\vspace{10pt}


\begin{table}[H]
{\tiny
\begin{tabular}{|c|c|c|c|c|c|}
\hline
 $h=0$ &  $h= \tfrac12$  & $h=1$ & $h=\tfrac32$ & $h=2$  & $h=\tfrac52$
            \\ \hline
SGRAV$_3$ $\left( \tfrac92 \right)_3$ & LGRAV$_3$ $\left( \tfrac{3\sqrt{6}}{2} \right)_{\frac{1}{2}}$ , $\left( \sqrt{15} \right)_{\frac{3}{2}}$ , $\left( \tfrac{\sqrt{70}}{2} \right)_{\frac{5}{2}}$  & LGRAV$_3$ $\left( \tfrac{3\sqrt{5}}{2} \right)_0$ , $\left( \tfrac72 \right)_1$ , $\left( \tfrac{\sqrt{57}}{2} \right)_2$ & LGRAV$_3$ $\left( \tfrac{\sqrt{42}}{2} \right)_{\frac{3}{2}}$  & SGINO$_3$ $\left( 3 \right)_2$  & SVEC$_3$ $\left( \tfrac52 \right)_{\frac{5}{2}}$ \\[4pt]
LGRAV$_3$ $\left( \tfrac{\sqrt{61}}{2} \right)_1$ & LGINO$_3$ $2 \times \left( \tfrac{\sqrt{57}}{2} \right)_{\frac{1}{2}}$ , $\left( \tfrac{3\sqrt{7}}{2} \right)_{\frac{3}{2}}$ , $\left( \tfrac{\sqrt{73}}{2} \right)_{\frac{5}{2}}$ & LGINO$_3$ $\left( 2\sqrt{3} \right)_0$ , $\left( \sqrt{13} \right)_1$ , $2\times \left( \sqrt{15} \right)_2$ & LGINO$_3$ $\left( \tfrac{\sqrt{39}}{2} \right)_{\frac{1}{2}}$ , $2 \times \left( \tfrac{3\sqrt{5}}{2} \right)_{\frac{3}{2}}$   & LGINO$_3$ $\left( \sqrt{7} \right)_1 $ &   \\
LGINO$_3$ $2 \times \left( 4 \right)_1$ , $\left( \sqrt{21} \right)_3$   &   &   &  &  & \\[6pt]
 \hline
\end{tabular}
}\normalsize
\caption{\footnotesize{$\cN=3$ supermultiplets at KK level $k=3$}\normalsize}
\label{tab:N=3multipletsatlevel3}
\end{table}

\end{landscape}


\newpage

 \begin{table}[H]

\centering

\scriptsize{

\begin{tabular}{llcccccl}
\hline
\multicolumn{2}{c}{$\Delta$}  && $k $  &&   $\textrm{SO}(3)_\cR \times \textrm{SO}(3)_\cF $  && $\cN=3$ supermultiplet
            \\ \hline
$1$ & $1.00$ & &  0   &&  $ \left(1,1 \right)  $   && MVEC$_3$ $\left( 1 \right)_1  $ \\[4pt]
%
 %
 %
$\tfrac32$ & $1.50$ & &  0   &&  $ \left( \tfrac12 ,\tfrac12 \right)  $   && SGINO$_3$ $\left( \tfrac32 \right)_{\tfrac12 } $ \\[4pt]
%
 %
 %
$\tfrac32$ & $1.50$ & &  1   &&  $ \left( \tfrac32 ,\tfrac32 \right)  $   && SVEC$_3$ $\left( \tfrac32 \right)_{\tfrac32 } $ \\[4pt]
%
%
 %
 %
$\sqrt{3}$ & $1.73$ & &  0   &&  $ \left( 0 , 0 \right)  $   && LGINO$_3$ $\left( \sqrt{3} \right)_{ 0 } $ \\[4pt]
%
%
 %
 %
$\sqrt{3}$ & $1.73$ & &  1   &&  $ \left( 0 , 1 \right)  $   && LGINO$_3$ $\left( \sqrt{3} \right)_{ 0 } $ \\[4pt]
%
 %
 %
 %
$2$ & $2.00$ & &  0   &&  $\left( 1 ,1 \right)  $   && MVEC$_3$ $\left( 1 \right)_1  $ \\[4pt]
%
%
 %
 %
$2$ & $2.00$ & & 1   &&  $ \left( 1 , 1 \right)  $   && SGINO$_3$ $\left( 2 \right)_{1 } $ \\[4pt]
%
%
 %
 %
$2$ & $2.00$ & & 2   &&  $ \left( 2 , 2 \right)  $   && SVEC$_3$ $\left( 2 \right)_{2 } $ \\[4pt]
%
%
%
 %
 %
$\tfrac{\sqrt{19}}{2}$ & $2.18$ & &  2   &&  $ \left( \tfrac12 , \tfrac32 \right)  $   && LGINO$_3$ $ \left( \tfrac{\sqrt{19}}{2} \right)_{ \tfrac12 } $ \\[4pt]
%
%
%
%
 %
 %
$\tfrac{\sqrt{21}}{2}$ & $2.29$ & &  1   &&  $ \left( \tfrac12 , \tfrac12 \right)  $   && LGINO$_3$ $ \left( \tfrac{\sqrt{21}}{2} \right)_{ \tfrac12 } $ \\[4pt]
%
%
 %
 %
$\tfrac52$ & $2.50$ & &  0   &&  $ \left( \tfrac32 ,\tfrac12 \right)  $   && SGINO$_3$ $\left( \tfrac32 \right)_{\tfrac12 } $ \\[4pt]
%
%
 %
 %
$\tfrac52$ & $2.50$ & &  0   &&  $ \left( \tfrac12 ,\tfrac12 \right)  $   && SGINO$_3$ $\left( \tfrac32 \right)_{\tfrac12 } $ \\[4pt]
%
%
 %
 %
$\tfrac52$ & $2.50$ & &  1   &&  $ \left( \tfrac32 ,\tfrac32 \right)  $   && SVEC$_3$ $\left( \tfrac32 \right)_{\tfrac32 } $ \\[4pt]
%
%
 %
 %
$\tfrac52$ & $2.50$ & &  1   &&  $ \left( \tfrac12 ,\tfrac32 \right)  $   && SVEC$_3$ $\left( \tfrac32 \right)_{\tfrac32 } $ \\[4pt]
%
%
%
 %
 %
$\tfrac52$ & $2.50$ & &  2   &&  $ \left( \tfrac32 ,\tfrac32 \right)  $   && SGINO$_3$ $\left( \tfrac52 \right)_{\tfrac32 } $ \\[4pt]
%
%
%
 %
 %
$\tfrac52$ & $2.50$ & &  3   &&  $ \left( \tfrac52 ,\tfrac52 \right)  $   && SVEC$_3$ $\left( \tfrac52 \right)_{\tfrac52 } $ \\[4pt]
%
%
 %
 %
$\tfrac12 + \tfrac{3\sqrt{2}}{2}$ & $2.62$ & &  1   &&  $ \left( \tfrac32 , \tfrac12 \right)  $   && LGRAV$_3$ $ \left( \tfrac{3\sqrt{2}}{2} \right)_{ \tfrac12 } $ \\[4pt]
%
%
%
 %
 %
$\tfrac12 + \tfrac{3\sqrt{2}}{2}$ & $2.62$ & &  1   &&  $ \left( \tfrac12 , \tfrac12 \right)  $   && LGRAV$_3$ $ \left( \tfrac{3\sqrt{2}}{2} \right)_{ \tfrac12 } $ \\[4pt]
%
%
%
 %
 %
$\sqrt{7}$ & $2.64$ & &  1   &&  $ \left( 1 , 0 \right)  $   && LGINO$_3$ $\left( \sqrt{7} \right)_{ 1 } $ \\[4pt]
%
%
%
 %
 %
$\sqrt{7}$ & $2.64$ & &  3   &&  $ \left( 1 , 2 \right)  $   && LGINO$_3$ $\left( \sqrt{7} \right)_{ 1 } $ \\[4pt]
%
%
 %
 %
$1+ \sqrt{3}$ & $2.73$ & &  0   &&  $ \left( 2 , 0 \right)  $   && LGINO$_3$ $\left( \sqrt{3} \right)_{ 0 } $ \\[4pt]
%
%
 %
 %
$1+ \sqrt{3}$ & $2.73$ & &  0   &&  $ \left( 0 , 0 \right)  $   && LGINO$_3$ $\left( \sqrt{3} \right)_{ 0 } $ \\[4pt]
%
%
 %
 %
$1+ \sqrt{3}$ & $2.73$ & &  1   &&  $ \left( 2 , 1 \right)  $   && LGINO$_3$ $\left( \sqrt{3} \right)_{ 0 } $ \\[4pt]
%
%
 %
 %
$1+ \sqrt{3}$ & $2.73$ & &  1   &&  $ \left( 0 , 1 \right)  $   && LGINO$_3$ $\left( \sqrt{3} \right)_{ 0 } $ \\[4pt]
%
 %
 %
$2\sqrt{2}$ & $2.83$ & &  2   &&  $ 2\times \left( 1 , 1 \right)  $   && LGINO$_3$ $2 \times \left( 2\sqrt{2} \right)_{ 1 } $ \\[4pt]
%
 %
  %
 %
$3$ & $3.00$ & & 1   &&  $ \left( 1 , 0 \right)  $   && SGRAV$_3$ $\left( \tfrac52 \right)_{1 } $ \\[4pt]
%
 %
  %
 %
$3$ & $3.00$ & & 1   &&  $ \left( 0 , 0 \right)  $   && SGRAV$_3$ $\left( \tfrac52 \right)_{1 } $ \\[4pt]
%
 %
  %
 %
$3$ & $3.00$ & & 1   &&  $ \left( 2 , 1 \right)  $   && SGINO$_3$ $\left( 2 \right)_{1 } $ \\[4pt]
%
%
 %
  %
 %
$3$ & $3.00$ & & 1   &&  $ 2 \times \left( 1 , 1 \right)  $   && SGINO$_3$ $\left( 2 \right)_{1 } $ \\[4pt]
%
%
 %
 %
$3$ & $3.00$ & & 2   &&  $ \left( 2 , 2 \right)  $   && SVEC$_3$ $\left( 2 \right)_{2 } $ \\[4pt]
%
%
 %
 %
$3$ & $3.00$ & & 2   &&  $ \left( 1 , 2 \right)  $   && SVEC$_3$ $\left( 2 \right)_{2 } $ \\[4pt]
%
%
 %
 %
$3$ & $3.00$ & & 2   &&  $ \left( 0 , 2 \right)  $   && SVEC$_3$ $\left( 2 \right)_{2 } $ \\[4pt]
%
%
 %
 %
$3$ & $3.00$ & & 3   &&  $ \left( 2 , 2 \right)  $   && SGINO$_3$ $\left( 3 \right)_{2 } $ \\[4pt]
%
%
 %
 %
$3$ & $3.00$ & & 4   &&  $ \left( 3 , 3 \right)  $   && SVEC$_3$ $\left( 3 \right)_{3 } $ \\[4pt]
%
%
%
%
 \hline
\end{tabular}

}\normalsize

\caption{\scriptsize{All KK scalars with dimension $\Delta \leq 3$ around the $\cN=3$ AdS$_4$ type IIA solution. }\normalsize}
\label{tab:N=3ISO7Scalars}
\end{table}

The spectrum of KK scalars  about the $\cN=3$ solution follows from the above results. Table \ref{tab:N=3ISO7Scalars} compiles all the scalars with conformal dimension $\Delta \leq 3$. For each scalar, the analytical value of $\Delta$, together with a convenient numerical approximation, is shown. The table also specifies the KK level $k$ and $\textrm{SO}(3)_\cR \times \textrm{SO}(3)_\cF $ quantum numbers $(r,h)$ of each scalar, together with the OSp$(4|3)$ supermultiplet where it belongs. The supermultiplets are labelled with their dimension and isospin $(E_0)_j$ (which may or may not coincide with those, $\Delta$ and $r$, for each scalar), and arise at the same KK level $k$ and have the same flavour charge $h$ than the scalars they contain. The OSp$(4|3)$ representation contents tabulated in appendix \ref{sec:N=3Supermultiplets} are useful to determine the scalar charges and dimensions reported in table \ref{tab:N=3ISO7Scalars}. All scalars dual to relevant and classically marginal operators of the dual field theory arise at levels $k \leq 4$. Each of these  KK levels also contain scalars dual to irrelevant operators. All scalars at KK levels $k \geq 5$ are dual to irrelevant operators.


\section{Discussion} \label{sec:Discussion}


In this paper, I have derived the KK vector mass matrix of a class of AdS$_4$ solutions of $D=11$ supergravity and massive type IIA supergravity from E$_{7(7)}$ ExFT \cite{Hohm:2013pua,Hohm:2013uia,Ciceri:2016dmd}, following \cite{Malek:2019eaz}. Then, I have used these and previous partial results \cite{Klebanov:2008vq,Klebanov:2009kp,Pang:2017omp,Pang:2015rwd} to determine the complete supersymmetric KK spectrum of some $\cN=2$ and $\cN=3$ solutions in this class. The $\cN=2$ AdS$_4$ CPW solution in $D=11$ \cite{Corrado:2001nv} is dual to an $\cN=2$ Chern-Simons CFT with gauge group $\textrm{U}(N) \times \textrm{U}(N)$. This CFT is defined on a stack of M2-branes, and arises as the infrared fixed point of a superpotential mass deformation \cite{Klebanov:2008vq} of ABJM \cite{Aharony:2008ug}. The $\cN=2$ and $\cN=3$ CFTs dual to the AdS$_4$ solutions \cite{Guarino:2015jca,Pang:2015vna,DeLuca:2018buk} of massive type IIA arise instead as superconformal fixed points of the D2-brane field theory, three-dimensional $\cN=8$ SU$(N)$ super-Yang Mills, augmented with Chern-Simons terms \cite{Guarino:2016ynd,Guarino:2019snw}. These CFTs have been described in \cite{Guarino:2015jca}. 

The present results determine the complete spectrum of single-trace operators with dimension of order one for these CFTs, in the strong coupling regime and at large $N$. See \cite{Gabella:2012rc,Fluder:2015eoa,Araujo:2016jlx} for some results on the large-$N$ spectrum of operators with dimensions that scale with $N$ raised to various powers, for some of these CFTs. The $\cN=2$ CFTs discussed above are intrinsically strongly coupled. In contrast, the $\cN=3$ CFT admits a weakly coupled limit which has been investigated in \cite{Gaiotto:2007qi,Minwalla:2011ma}. The spectrum of operators of the $\cN=3$ CFT that lie in short representations of OSp$(4|3)$ has been computed at weak coupling and large $N$ directly from the field theory \cite{Minwalla:2011ma}. This short spectrum can be expected to be subject to non-renormalisation theorems and, for this reason, both its structure in terms of $\textrm{OSp}(4|3) \times \textrm{SO}(3)_\cF$ representations and the conformal dimensions $E_0$ of these operators must remain unaltered at strong coupling. Satisfactorily, the bulk calculation of the short spectrum at strong coupling found in table \ref{tab:ShortN=3Spectrum} perfectly matches the short spectrum reported at weak coupling in table 15 of  \cite{Minwalla:2011ma}.

The analysis in this paper determines, in particular, the spectrum of relevant and classically marginal operators of these CFTs. For example, the $\cN=2$ CFT dual to the massive type IIA AdS$_4$ solution \cite{Guarino:2015jca} contains SU$(N)$-adjoint hypermultiplets ${\cal Z}^a$, $a=1,2,3$, in the fundamental of the SU(3) flavour group. The theory has a superpotential, ${\cal W} \sim \textrm{tr} \,  \epsilon_{abc}  \, {\cal Z}^a  [ {\cal Z}^b ,  {\cal Z}^c ] $, analogous to that of four-dimensional $\cN=4$ super-Yang-Mills written out in $\cN=1$ language. Tables \ref{tab:ShortN=2Spectrum} and \ref{tab:multipletsatlevel0} (see also table 5 of 
\cite{Guarino:2019snw}) show the existence of $\bm{6}$ hypermultiplets $\textrm{tr} \, {\cal Z}^{(a} {\cal Z}^{b)}$, which arise at KK level $k=0$. These are superpotential mass terms, thus relevant in agreement with table \ref{tab:N=2ISO7Scalars}, that can be added to the $\cN=2$ theory to generate renormalisation group (RG) flow \cite{Gaiotto:2007qi}. Being generated by a $k=0$ deformation, this flow can be holographically built in gauged supergravity \cite{Guarino:2019snw} in analogy with similar mass deformation flows of $\cN=4$ super-Yang-Mills \cite{Freedman:1999gp} and ABJM \cite{Ahn:2000aq,Corrado:2001nv}. 

Tables \ref{tab:ShortN=2Spectrum} and \ref{tab:multipletsatlevel1} also show the existence of $\bm{10}$ hypermultiplets of the form $\textrm{tr} \, {\cal Z}^{(a} {\cal Z}^b {\cal Z}^{c)}$  at KK level $k=1$ in the spectrum of the $\cN=2$ CFT with AdS$_4$ dual in massive IIA. These cubic terms can again be added to the superpotential to generate deformations that have direct analogues in four-dimensional $\cN=4$ super-Yang Mills and ABJM. In the ABJM case, this cubic superpotential deformation is relevant \cite{Jafferis:2011zi,Gabella:2012rc} (see also \cite{Cesaro:2020piw}) and generates RG flow. In the present massive type IIA case, the cubic deformation is instead classically marginal according to tables \ref{tab:ShortN=2Spectrum}, \ref{tab:multipletsatlevel1} and \ref{tab:N=2ISO7Scalars}. This is exactly like the analogue $\cN=1$ deformation of $\cN=4$ super-Yang-Mills. As is well known, out of the classically marginal $\cN=1$ deformations of $\cN=4$ super-Yang-Mills, only the so-called $\beta$- and cubic deformations are exactly marginal. It would be interesting to determine the exactly marginal deformations in the $\cN=2$ massive IIA case, and engineer their gravity duals following \cite{Lunin:2005jy,Ashmore:2016oug}. The exactly marginal deformations of the $\cN=3$ theory that preserve $\cN=2$ have been determined in \cite{Minwalla:2011ma} at weak coupling. These are expected to be non-renormalised, and should thus be also included in table \ref{tab:N=3ISO7Scalars} above.

On a different note, it was discussed in \cite{Pang:2017omp,Dimmitt:2019qla} that the KK graviton mass matrix for the $D=11$ and massive type IIA AdS$_4$ solutions that uplift from SO(8) and ISO(7) supergravity displays some universality behaviour. Specifically, the graviton traces match for solutions in both gaugings with the same residual (super)symmetry, at fixed SO(8) KK level $n$ and combined SO(7) KK levels $k = 0 , 1 , \ldots , n$ (in order to trace over the same number of states, through $[n,0,0,0] \rightarrow \oplus_{k=0}^n [k,0,0]$). I have checked that the same holds for the KK vector mass matrix (\ref{eq:KKVecMassMat}) with (\ref{eq:MassAux1}), (\ref{eq:MassAux2}). However, the traces now involve the unphysical states that are eaten by massive gravitons, discussed in section \ref{sec:KKVecMassMatCalc}. Thus, these mass matrix traces carry no physical significance. Matchings still occur for some solutions when tracing over physical states only. For example, the vector mass matrix trace taken over physical states for the SO$(7)_v$ and SO$(7)_c$ AdS$_4$ solutions of $D=11$ supergravity match. Also, while they do not share the same symmetry, there is a similar match tracing over physical KK vectors states for the $\cN=0$ G$_2$ and $\cN=3$ SO(4) AdS$_4$ solutions in massive IIA. These solutions have the same cosmological constant \cite{Guarino:2015qaa}, and the KK graviton traces also match \cite{Pang:2017omp}.

The SO(8) and ISO(7) gaugings considered in this paper also have interesting $\cN=1$, or even non-supersymmetric, AdS$_4$ solutions. In these cases with low or no supersymmetry, the KK scalar spectra is not implied by the KK graviton and vector spectra, and will need to be computed independently. The scalar spectrum of some $D=11$ and type IIB AdS solutions have already been computed using ExFT techniques in \cite{Malek:2019eaz,Malek:2020mlk}.


\section*{Acknowledgements}


Praxitelis Ntokos and Junchen Rong are kindly thanked for discussions. This work was supported by the NSF grant PHY-2014163 and, partially, by grants SEV-2016-0597 and PGC2018-095976-B-C21 from MCIU/AEI/FEDER, UE.


\appendix


\section{KK vector spectra of selected AdS$_4$ solutions} \label{sec:SpecificSpectra}

This paper follows the conventions for $D=11$ \cite{Cremmer:1978km} and $D=4$ $\cN=8$ SO(8) \cite{deWit:1982ig} supergravity specified in \cite{Varela:2015ywx}. The conventions for massive IIA \cite{Romans:1985tz} and $D=4$ $\cN=8$ ISO(7) supergravity are spelled out in \cite{Guarino:2015vca} and \cite{Guarino:2015qaa}, respectively. The generic conventions for $D=4$ $\cN=8$ gauged supergravity \cite{deWit:2007mt}, including the E$_{7(7)}$ generators, are those of \cite{Guarino:2015vca}.

The KK vector mass matrix $({\cal M}^2)_{\underline{M} \Lambda}{}^{\underline{N} \Sigma}$,  (\ref{eq:KKVecMassMat}) with (\ref{eq:MassAux1}), (\ref{eq:MassAux2}), for the type of AdS$_4$ solutions of $D=11$ and type IIA under consideration depends exclusively on data of $D=4$ $\cN=8$ supergravity, as well as on the SO(8) or SO(7) generators. The latter are left as footprints of the topological $S^7$ and $S^6$ that these solutions feature as their internal spaces. All the tensors that enter (\ref{eq:KKVecMassMat}) are constant, except for the matrix $M_{\underline{M} \underline{N}} $ (and its inverse $M^{\underline{M} \underline{N}} $), which is quadratic, $M = {\cal V} {\cal V}^{\textrm{T}}$, in the $\textrm{E}_{7(7)}/\textrm{SU}(8)$ coset representative ${\cal V} (\varphi)$  that contains the $D=4$ $\cN=8$ scalars $\varphi$. When evaluated at a critical point of the $D=4$ $\cN=8$ potential $V$, $\varphi_*$ such that $\partial V (\varphi_*) = 0 $, the matrices $M_{\underline{M} \underline{N}} (\varphi_*)$ and $M^{\underline{M} \underline{N}} (\varphi_*)$ also become constant, of course, though still solution-dependent.

The differences between $D=11$ and type IIA solutions are introduced in (\ref{eq:KKVecMassMat}) by the embedding tensor $\Theta_{\underline{M}}{}^\alpha$ and the generators $({\cal T}_{\underline{M} })_\Lambda{}^\Sigma$. For both $D=4$ $\cN=8$ gaugings considered, the embedding tensor can be specified as follows. Introducing indices $A=1 , \ldots , 8$, in the fundamental, $\bm{8}$, of SL(8) (and in the $\bm{8}_v$ of SO(8), as in (\ref{eq:SigmaIndex})) the non-vanishing components of the embedding tensor $\Theta_{\underline{M}}{}^\alpha = \left( \Theta_{\underline{M}}{}^{C}{}_D \; , \; \Theta_{\underline{M}}{}^{CDEF} \equiv 0 \right)$ are $\Theta_{\underline{M}}{}^{C}{}_D = \left( \Theta_{AB}{}^{C}{}_D \; , \; \Theta^{ABC}{}_D \right)$, where \cite{Dall'Agata:2012bb}
\begin{equation} \label{eq:SO8EmbTen}
\Theta_{AB}{}^C{}_D = 2 \,  \delta^C_{[A} \theta_{B]D} \; , \qquad 
\Theta^{ABC}{}_D =  2 \, \delta^{[A}_D \xi^{B]C} \; ,
\end{equation}
with
\begin{eqnarray} \label{eq:thetaSO8}
\textrm{SO(8) gauging} & : & \theta_{AB} = \delta_{AB} \; , \qquad \qquad \qquad  \qquad  \quad \; \; \xi^{AB} = 0 \; ,  \\[5pt]
\label{eq:thetaISO7}
\textrm{ISO(7) gauging} & : & \theta_{AB} = \textrm{diag} \, \left( 1,1,1,1,1,1,1,0 \right)   \; , \quad \xi^{AB} =  \left( 0,0,0,0,0,0,0,1 \right)   . \hspace{10pt}
\end{eqnarray}
As for the tensors $({\cal T}_{\underline{M} })_\Lambda{}^\Sigma = \left( ({\cal T}_{AB})_\Lambda{}^\Sigma \; , \; ({\cal T}^{AB})_\Lambda{}^\Sigma \equiv 0 \right) $, depending on how they are defined they encode the SO(8) or SO(7) generators in the infinite-dimensional, reducible representations (\ref{eq:SymTrac}). As discussed in section \ref{sec:KKVecMassMatCalc}, the mass matrix (\ref{eq:KKVecMassMat}) is block diagonal, and can be diagonalised KK level by KK level. This implies that each individual block can be treated independently. One can then focus on the generators in just the symmetric traceless representations of SO(8) and SO(7). More concretely, breaking out the $\Lambda$, $\Sigma$ indices as in (\ref{eq:SigmaIndex}), one has
\begin{eqnarray} \label{eq:GenSOSymTrSO8}
\textrm{SO}(8) &: & ({\cal T}_{AB})_{C_1 \ldots C_n}{}^{D_1 \ldots D_n} = n \, (  {\cal T}_{AB})_{\{ C_1}{}^{ \{ D_1} \delta_{C_2}^{D_2} \ldots    \delta_{C_n\}}^{D_n\}} \; ,  \\[5pt]
\label{eq:GenSOSymTrISO7}
\textrm{SO}(7) &: & ({\cal T}_{AB})_{K_1 \ldots K_k}{}^{L_1 \ldots L_k}  \quad =   k \, (  {\cal T}_{AB})_{\{ K_1}{}^{ \{ L_1} \delta_{K_2}^{L_2} \ldots    \delta_{K_k\}}^{L_k\}} \; .
\end{eqnarray}
In (\ref{eq:GenSOSymTrSO8}), $( {\cal T}_{AB})_{C}{}^{ D} \equiv 2 \delta_{[A}^{D} \theta_{B]C}$, with $\theta_{AB}$ given in (\ref{eq:thetaSO8}), are the $\bm{28}$ generators of SO(8) in the fundamental representation. In (\ref{eq:GenSOSymTrISO7}), the tensor $( {\cal T}_{AB})_{K}{}^{ L} \equiv 2 \delta_{[A}^{L} \theta_{B]K}$ is defined using $\theta_{AB}$ in (\ref{eq:thetaISO7}), with the index splitting $A = (I, 8)$, $I=1, \ldots , 7$. It contains the $\bm{21}$ generators  $( {\cal T}_{IJ})_{K}{}^{ L} $ of SO(7), together with $( {\cal T}_{I8})_{K}{}^{ L} =0$. 

I have used these formulae to compute the first few levels of the KK vector spectrum of some AdS$_4$ solutions of $D=11$ and massive IIA supergravity that uplift from critical points of the corresponding $D=4$ $\cN=8$ supergravities. The KK vector spectra of all the $D=10, 11$ solutions in this class can be computed using this technology, even if the solutions are only known as critical points of the $D=4$ supergravity and the higher-dimensional uplift is not explicitly known. The latest word on the classification of critical points of $D=4$ $\cN=8$ SO(8) supergravity is \cite{Comsa:2019rcz}. For $D=4$ $\cN=8$ ISO(7) supergravity, the latest classification of critical points can be found in \cite{Guarino:2020jwv}, although the latter should admit improvements from the machine learning methods employed in \cite{Comsa:2019rcz}. 

For definiteness, I have focused on the $D=11$ solutions that preserve at least an SU(3) subgroup of SO(8), following the conventions of \cite{Larios:2019kbw}. The $D=4$ critical points in this sector were classified in \cite{Warner:1983vz}, and their $D=11$ uplift is also known \cite{Freund:1980xh,Corrado:2001nv,deWit:1984nz,deWit:1984va,Englert:1982vs,Pope:1984bd}. The entire KK spectrum of the $\cN=8$ SO(8)-invariant solution \cite{Freund:1980xh} has long been known \cite{Englert:1983rn,Sezgin:1983ik,Biran:1983iy}, and my results reproduce their KK vector spectrum. For the $\cN=2$ $\textrm{SU}(3) \times \textrm{U}(1)$ solution \cite{Warner:1983vz,Corrado:2001nv}, the spectrum for all KK fields at levels $n=0$ \cite{Nicolai:1985hs} and $n=1$ \cite{Malek:2019eaz} is known, along with the entire KK graviton spectrum \cite{Klebanov:2009kp}. My results for the KK vectors again agree with \cite{Nicolai:1985hs,Malek:2019eaz} and extend them to higher KK levels. The KK graviton spectra of the other solutions has been computed in \cite{Dimmitt:2019qla}, and the KK vector spectra reported below are new. The results are summarised for levels $n=0,1,2$ in table \ref{tab:KKVectorsSO8}. In the table, the mass $M^2$ eigenvalues have been normalised to the corresponding AdS$_4$ radius squared, $L^2 = -6/V$, where $V <0$ is the cosmological constant of that point. The eigenvalues are given as $ (M^2 L^2)^{(p)}$, where $p$ is a positive integer that denotes the multiplicity. Recall that the scaling dimension $\Delta$ of a vector of mass $ M^2 L^2$ is given by 
\begin{equation}
(\Delta-1) (\Delta-2) = M^2 L^2 \; .
\end{equation}
This formula has been used throughout to convert the KK vector mass eigenvalues to the conformal dimensions reported in the main text. For completeness, recall that for gravitons and scalars the analogue relation is
\begin{equation}
\Delta (\Delta-3) = M^2 L^2 \; .
\end{equation}

Turning to massive IIA, I have again focused for concreteness on the solutions that preserve the SU(3) subgroup of SO(7). These solutions were classified as critical points of $D=4$ $\cN=8$ ISO(7) supergravity in \cite{Guarino:2015qaa}. The latter reference also contains the spectrum for all bosonic fields at KK level $k=0$. Due to its particular interest, I have also looked at the $\cN=3$ solution with SO(4) residual symmetry, which lies outside of this class. This solution was first found as a critical point of the $D=4$ supergravity in \cite{Gallerati:2014xra}, where the $k=0$ KK spectrum was also determined. The type IIA uplift of all these solutions is known \cite{Guarino:2015jca,Varela:2015uca,Pang:2015vna,DeLuca:2018buk}. The KK spectrum of gravitons at all levels is also known \cite{Pang:2015rwd,Pang:2017omp}. The present results reproduce the vector spectrum at $k=0$ level and extend them to higher KK levels. The results for levels $k=0,1,2$ are summarised in table \ref{tab:KKVectorsISO7}. The format is the same as table \ref{tab:KKVectorsSO8}.


\newpage

\begin{landscape}

 \begin{table}[H]
\centering


\scriptsize{

\begin{tabular}{|l|l|l|}
\hline
Sol.                & $n$                  & $ M^2 L^2$                                     \\ \hline \hline
\multirow{4}{*}{$\cN=8$, SO(8)} & $0$    & $0^{(28)}$                                \\[4pt]  
\cline{2-3} 
  & 1 & $\tfrac{15}{4}^{(56)}$ , $\tfrac{3}{4}^{(160)}$          \\[4pt]  
\cline{2-3} 
& $ 2$ & $12^{(28)}$ , $6^{(350)}$ , $2^{(567)}$                              \\[4pt]  \hline \hline
\multirow{6}{*}{$\cN=2$, U(3)} & $0$    & $4^{(1)}$ ,   $\tfrac{28}{9}^{(6)}$ ,   $\tfrac{4}{9}^{(12)}$   , $0^{(9)}$                             \\[4pt]  
\cline{2-3} 
  & 1 &   $\left( \tfrac{43}{9} \pm \tfrac{\sqrt{145}}{3} \right)^{(6)}$ ,    $8^{(2)}$ ,   $\left( \tfrac{70}{9} \right)^{(12)}$ , $6^{(4)}$ , $\left( \tfrac{52}{9}  \right)^{(6)}$ ,  $\left( 3 \pm \sqrt{3} \right) ^{(32)}$  ,  $\tfrac{40}{9}^{(12)}$ ,  $\tfrac{34}{9}^{(24)}$ ,  $\left( \tfrac{28}{9} \right)^{(24)}$ , $2^{(2)}$ , $\tfrac{10}{9}^{(24)}$ , $\tfrac{4}{9}^{(30)}$  \\[4pt]  
\cline{2-3} 
& \multirow{2.8}{*}{$2$} &        $\left( 11 \pm \sqrt{41} \right)^{(1)}$  , $\left( \tfrac{91}{9} \pm \tfrac{\sqrt{337}}{3} \right)^{(6)}$ , $\left( \tfrac{85}{9} \pm \tfrac{\sqrt{313}}{3} \right)^{(6)}$ , $\tfrac{130}{9}^{(12)}$ , $14^{(2)}$ , $\left( \tfrac{88}{9} \pm \tfrac{\sqrt{88}}{3} \right)^{(24)}$ , $12^{(14)}$ , $\left( \tfrac{61}{9} \pm \tfrac{\sqrt{217}}{3} \right)^{(12)}$ , $\tfrac{100}{9}^{(6)}$ , $\left(  8 \pm \sqrt{8} \right)^{(64)}$ ,
\\[4pt] 
\cline{3-3} 
&  &     $\tfrac{94}{9}^{(6)}$ ,  $10^{(4)}$ ,      $\tfrac{88}{9}^{(24)}$ ,       $\tfrac{82}{9}^{(24)}$ ,        $\tfrac{76}{9}^{(36)}$ ,   $8^{(16)}$ ,       $\tfrac{70}{9}^{(36)}$ ,  $\left( \tfrac{40}{9} \pm \tfrac{2\sqrt{10}}{3} \right)^{(120)}$   ,  $6^{(54)}$ ,    $\tfrac{52}{9}^{(48)}$   ,    $\tfrac{46}{9}^{(30)}$  ,     $\tfrac{40}{9}^{(24)}$  ,    $4^{(20)}$  ,  $2^{(48)}$   ,  $\tfrac{4}{3}^{(27)}$   ,  $\tfrac{10}{9}^{(48)}$        \\[4pt]  \hline \hline
\multirow{4.5}{*}{$\cN=1$, G$_2$} & $0$    & $\left( \tfrac{3}{2} \pm \sqrt{\tfrac{3}{2}}  \right)^{(7)}$    , $0^{(14)}$                            \\[4pt]  
\cline{2-3} 
  & 1 & $\left( \tfrac{47}{8} \pm \sqrt{\tfrac{47}{8}}  \right)^{(7)}$ ,   $\tfrac{55}{8}^{(7)} $   ,   $\tfrac{51}{8}^{(1)} $   , $\left( \tfrac{27}{8} \pm \sqrt{\tfrac{27}{8}}  \right)^{(14)}$ ,   $\tfrac{35}{8}^{(14)} $  , $\left( \tfrac{61}{24} \pm \sqrt{\tfrac{61}{24}}  \right)^{(27)}$ , $\tfrac{31}{8}^{(7)} $ , $\tfrac{85}{24}^{(27)} $ , $\tfrac{5}{8}^{(64)} $  \\[4pt]  
\cline{2-3} 
& $ 2$ & $\left( \tfrac{23}{2} \pm \sqrt{\tfrac{23}{2}}  \right)^{(14)}$    , $\tfrac{25}{2}^{(14)} $  , $12^{(29)}$  , $\left( \tfrac{49}{6} \pm \sqrt{\tfrac{49}{6}}  \right)^{(27)}$     , $10^{(28)}$   ,  $\tfrac{19}{2}^{(7)} $ , $\tfrac{55}{6}^{(54)} $ , $\left( \tfrac{21}{4} \pm \sqrt{\tfrac{21}{4}}  \right)^{(64)}$ ,  $\tfrac{25}{4}^{(128)} $ ,  $\tfrac{37}{6}^{(27)} $  $6^{(105)}$ , $5^{(77)}$, $2^{(77)}$ ,  $\tfrac{5}{3}^{(189)} $  \\[4pt]  \hline \hline
\multirow{4}{*}{$\cN=0$, SO$(7)_v$} & $0$    & $\left( \tfrac{12}{5} \right)^{(7)}$    , $0^{(21)}$          \\[4pt]  
\cline{2-3} 
& $ 1$ &   $\left( \tfrac{153}{20} \right)^{(7)}$ , $\left( \tfrac{21}{4} \right)^{(42)}$ , $\left( \tfrac{81}{20} \right)^{(35)}$ , $\left( \tfrac{57}{20} \right)^{(27)}$ ,  $\left( \tfrac{9}{20} \right)^{(105)}$ , 
 \\[4pt]  
\cline{2-3} 
& $ 2$ &    $\left( \tfrac{72}{5} \right)^{(14)}$ , $12^{(63)}$ ,  $\left( \tfrac{54}{5} \right)^{(35)}$ ,  $\left( \tfrac{48}{5} \right)^{(27)}$ , $\left( \tfrac{36}{5} \right)^{(210)}$ ,   $6^{(189)}$ , $\left( \tfrac{18}{5} \right)^{(77)}$ , $\left( \tfrac{6}{5} \right)^{(330)}$
\\[4pt]  \hline \hline
\multirow{4}{*}{$\cN=0$, SO$(7)_c$} & $0$    & $\left( \tfrac{12}{5} \right)^{(7)}$    , $0^{(21)}$          \\[4pt]  
\cline{2-3} 
& $ 1$ &   $\left( \tfrac{81}{10} \right)^{(8)}$ , $\left( \tfrac{39}{10} \right)^{(96)}$ , $\left( \tfrac{9}{10} \right)^{(112)}$ 
 \\[4pt]  
\cline{2-3} 
& $ 2$ &    $\left( \tfrac{72}{5} \right)^{(7)}$ , $12^{(42)}$ ,  $\left( \tfrac{54}{5} \right)^{(35)}$ ,  $\left( \tfrac{36}{5} \right)^{(105)}$ ,  $6^{(378)}$ , $\left( \tfrac{12}{5} \right)^{(378)}$ 
\\[4pt]  \hline \hline
\multirow{4}{*}{$\cN=0$, SU$(4)_c$} & $0$    & $6^{(1)}$ ,  $\left( \tfrac{9}{4} \right)^{(12)}$    , $0^{(15)}$          \\[4pt]  
\cline{2-3} 
& $ 1$ &   $\left( \tfrac{135}{16} \right)^{(24)}$ , $\left( \tfrac{63}{16} \right)^{(120)}$ , $\left( \tfrac{15}{16} \right)^{(72)}$ 
 \\[4pt]  
\cline{2-3} 
& $ 2$ & $18^{(1)}$ ,    $\left( \tfrac{57}{4} \right)^{(24)}$ , $12^{(90)}$ ,  $\left( \tfrac{45}{4} \right)^{(60)}$ , $9^{(20)}$ ,  $\left( \tfrac{27}{4} \right)^{(256)}$ ,  $6^{(270)}$ , $3^{(84)}$ ,  $\left( \tfrac{9}{4} \right)^{(140)}$
\\[4pt]  \hline 
\end{tabular}


}\normalsize

\caption{\footnotesize{The KK vector spectra up to KK level $n=2$
about the $D=11$ AdS$_4$ solutions that uplift on $S^7$ from vacua of $D=4$ $\cN=8$ SO(8) gauged supergravity with at least SU(3) symmetry.}\normalsize}
\label{tab:KKVectorsSO8}
\end{table}

\end{landscape}


\newpage

\begin{landscape}

 \begin{table}[H]
\centering


\tiny{

\begin{tabular}{|l|l|l|}
\hline
Sol.                & $k$                  & $M^2 L^2$                                     \\ \hline \hline
\multirow{6}{*}{$\cN=3$, SO(4)} &$0$    &   $\left(  3 \pm \sqrt{3} \right)^{(3)} $  , $ \tfrac{15}{4}^{(4)}$ , 
$ \tfrac{3}{4}^{(12)}$       , $0^{(6)}$             \\[4pt] 
\cline{2-3} 
  & 1 &  $12^{(1)}$ ,  $\left(  7 \pm \sqrt{7} \right)^{(9)} $ , $\left(  \tfrac{21}{4} \pm 3\sqrt{2} \right)^{(12)} $ ,   $\left(  \tfrac{21}{4} \pm \sqrt{\tfrac{21}{4}} \right)^{(12)} $ , $6^{(24)}$  , $\left(  3 \pm \sqrt{3} \right)^{(9)} $ , $\tfrac{17}{4}^{(36)}$   ,   $2^{(36)}$ ,  $\tfrac{3}{4}^{(8)}$  \\[4pt]  
\cline{2-3} 
& \multirow{2.8}{*}{$2$} &    $20^{(3)}$ , $\left(  \tfrac{43}{4} \pm 2\sqrt{10} \right)^{(24)} $ , $\left(  13 \pm \sqrt{13} \right)^{(15)} $ , $\left(  10 \pm \sqrt{37} \right)^{(3)} $ , $\left(  \tfrac{37}{4} \pm \sqrt{34} \right)^{(12)} $ , $\left(  \tfrac{43}{4} \pm \sqrt{\tfrac{43}{4} } \right)^{(24)} $ , $\left(  8 \pm \sqrt{29} \right)^{(27)} $ , $\left(  10 \pm \sqrt{10} \right)^{(6)} $ , $\left(  \tfrac{37}{4} \pm \sqrt{\tfrac{37}{4} } \right)^{(12)} $ 
\\[4pt] 
\cline{3-3} 
&  &   $12^{(24)}$  ,   $\left(  8 \pm \sqrt{8} \right)^{(54)} $  ,  $\tfrac{39}{4}^{(72)}$ , $9^{(9)}$  ,  $\tfrac{35}{4}^{(24)}$ , $\tfrac{33}{4}^{(36)}$ ,  $7^{(81)}$  , $\left(  \tfrac{19}{4} \pm \sqrt{\tfrac{19}{4} } \right)^{(24)} $ , $6^{(15)}$ , $ \tfrac{15}{4}^{(48)}$ ,  $2^{(15)}$  \\[4pt]  \hline \hline
\multirow{6}{*}{$\cN=2$, U(3)} & $0$    & $4^{(1)}$ ,   $\tfrac{28}{9}^{(6)}$ ,   $\tfrac{4}{9}^{(12)}$   , $0^{(9)}$                             \\[4pt]  
\cline{2-3} 
  & 1 &    $12^{(1)}$ ,   $\left( \tfrac{52}{9} \pm \tfrac{2\sqrt{13}}{3} \right)^{(12)}$ ,    $8^{(1)}$ ,   $\left( \tfrac{58}{9} \right)^{(6)}$ , $6^{(20)}$ , $\left( \tfrac{28}{9} \pm \tfrac{2\sqrt{7}}{3} \right)^{(24)}$ ,  $\tfrac{40}{9}^{(12)}$  ,  $4^{(8)}$ ,  $2^{(33)}$ ,  $\left( \tfrac{10}{9} \right)^{(36)}$ \\[4pt]  
\cline{2-3} 
& \multirow{2.8}{*}{$2$} &        $\left( \tfrac{37}{3} \pm \sqrt{\tfrac{139}{3} } \right)^{(1)}$  , $\left( \tfrac{154}{9} \right)^{(6)}$ , $\left( \tfrac{130}{9} \right)^{(24)}$ ,    $\left( \tfrac{25}{3} \pm \sqrt{\tfrac{91}{3}} \right)^{(8)}$    , $\left( \tfrac{40}{3} \right)^{(2)}$   ,    $\left( \tfrac{28}{3} \pm \sqrt{\tfrac{28}{3}} \right)^{(32)}$  , $\left( \tfrac{106}{9} \right)^{(6)}$    ,    $\left( \tfrac{76}{9} \pm \tfrac{\sqrt{76}}{3} \right)^{(24)}$   , $\left( \tfrac{34}{3} \right)^{(4)}$ ,  $\left( \tfrac{88}{9} \right)^{(66)}$    ,  $\left( \tfrac{28}{3} \right)^{(16)}$    ,  
\\[4pt] 
\cline{3-3} 
&  &  $\left( \tfrac{82}{9} \right)^{(12)}$  ,   $\left( \tfrac{52}{9} \pm \tfrac{\sqrt{52}}{3} \right)^{(60)}$    ,   $\left( \tfrac{70}{9} \right)^{(48)}$     ,     $\left( \tfrac{16}{3} \pm \sqrt{\tfrac{16}{3} } \right)^{(40)}$      ,   $\left( \tfrac{22}{3} \right)^{(32)}$      ,   $\left( \tfrac{58}{9} \right)^{(30)}$          ,   $\left( \tfrac{40}{9} \right)^{(66)}$          ,   $\left( \tfrac{28}{9} \right)^{(60)}$          ,   $\left( \tfrac{8}{3} \right)^{(27)}$                                 \\[4pt]  \hline \hline
\multirow{4.5}{*}{$\cN=1$, G$_2$} & $0$    & $\left( \tfrac{3}{2} \pm \sqrt{\tfrac{3}{2}}  \right)^{(7)}$    , $0^{(14)}$                            \\[4pt]  
\cline{2-3} 
  & 1 &   $12^{(1)}$ ,  $\left( \tfrac{15}{2} \right)^{(7)}$ ,  $6^{(15)}$ ,   $\left( \tfrac{19}{6} \pm \sqrt{\tfrac{19}{6}}  \right)^{(27)}$  ,  $\left( \tfrac{9}{2} \right)^{(7)}$ ,  $\left( \tfrac{25}{6} \right)^{(27)}$ ,  $2^{(14)}$ ,  $\left( \tfrac{5}{4} \right)^{(64)}$  \\[4pt]  
\cline{2-3} 
& $ 2$ & $\left( \tfrac{79}{6} \pm \sqrt{\tfrac{79}{6}}  \right)^{(7)}$    , $\tfrac{85}{6}^{(7)} $ , $\left( \tfrac{32}{3} \pm \sqrt{\tfrac{32}{3}}  \right)^{(14)}$    ,  $\tfrac{35}{3}^{(14)} $    ,  $\tfrac{65}{6}^{(27)} $ , $\left( \tfrac{83}{12} \pm \sqrt{\tfrac{83}{12}}  \right)^{(64)}$  , $\left( \tfrac{17}{3} \pm \sqrt{\tfrac{17}{3}}  \right)^{(77)}$    ,  $\tfrac{95}{12}^{(64)} $   ,  $\tfrac{47}{6}^{(27)} $  ,  $\tfrac{20}{3}^{(77)} $   ,  $\tfrac{10}{3}^{(189)} $      \\[4pt]  \hline \hline
\multirow{5}{*}{$\cN=1$, SU(3)} & $0$    & $6^{(1)}$ ,   $\tfrac{28}{9}^{(6)}$ ,    $\tfrac{25}{9}^{(6)}$ ,   $2^{(1)}$ ,  $\tfrac{4}{9}^{(6)}$   , $0^{(8)}$                             \\[4pt]  
\cline{2-3} 
  & 1 &    $12^{(2)}$  , $10^{(2)}$  , $\left( \tfrac{61}{9} \pm \sqrt{\tfrac{61}{9}}  \right)^{(12)}$ ,   $\tfrac{70}{9}^{(18)}$ ,  $7^{(1)}$ ,  $6^{(18)}$  , $\left( \tfrac{31}{9} \pm \sqrt{\tfrac{31}{9}}  \right)^{(12)}$ ,  $5^{(24)}$ , $\tfrac{43}{9}^{(6)}$ , $\tfrac{40}{9}^{(24)}$ ,  $2^{(16)}$  , $\tfrac{10}{9}^{(30)}$        \\[4pt]  
\cline{2-3} 
& \multirow{2.8}{*}{$2$} &     $\left( \tfrac{47}{3} \pm \sqrt{\tfrac{47}{3}}  \right)^{(2)}$ , $\tfrac{154}{9}^{(24)}$ , $\tfrac{50}{3}^{(3)}$ , $\tfrac{130}{9}^{(36)}$ ,   $\left( \tfrac{32}{3} \pm \sqrt{\tfrac{32}{3}}  \right)^{(32)}$ , $\tfrac{41}{3}^{(1)}$ , $\left( \tfrac{91}{9} \pm \sqrt{\tfrac{91}{9}}  \right)^{(24)}$ , $\tfrac{35}{3}^{(56)}$ , $\tfrac{103}{9}^{(6)}$ , $\tfrac{100}{9}^{(48)}$ , $\tfrac{88}{9}^{(24)}$ , $\left( \tfrac{61}{9} \pm \sqrt{\tfrac{61}{9}}  \right)^{(60)}$ , $\tfrac{26}{3}^{(8)}$ ,
\\
\cline{3-3} 
&  &  $\tfrac{73}{9}^{(12)}$ ,    $\left( \tfrac{17}{3} \pm \sqrt{\tfrac{17}{3}}  \right)^{(20)}$ , $\tfrac{70}{9}^{(120)}$      , $\tfrac{20}{3}^{(40)}$ ,  $\tfrac{10}{3}^{(27)}$ ,  $\tfrac{25}{9}^{(48)}$     \\[4pt]  \hline \hline
\multirow{4}{*}{$\cN=0$, SO$(7)$} & $0$    & $\left( \tfrac{12}{5} \right)^{(7)}$    , $0^{(21)}$          \\[4pt]  
\cline{2-3} 
& $ 1$ &  $12^{(1)}$ ,  $6^{(21)}$ ,  $\left( \tfrac{24}{5} \right)^{(35)}$ , $\left( \tfrac{18}{5} \right)^{(27)}$ , $\left( \tfrac{6}{5} \right)^{(105)}$ 
 \\[4pt]  
\cline{2-3} 
& $ 2$ &    $\left( \tfrac{82}{5} \right)^{(7)}$ , $14^{(21)}$ ,  $\left( \tfrac{46}{5} \right)^{(105)}$ ,  $8^{(189)}$ , $\left( \tfrac{28}{5} \right)^{(77)}$ ,  $\left( \tfrac{16}{5} \right)^{(330)}$ 
\\[4pt]  \hline \hline
\multirow{4}{*}{$\cN=0$, SO$(6)$} & $0$    & $6^{(1)}$ ,  $\left( \tfrac{9}{4} \right)^{(12)}$    , $0^{(15)}$          \\[4pt]  
\cline{2-3} 
& $ 1$ & $12^{(2)}$ ,   $\left( \tfrac{33}{4} \right)^{(18)}$ ,  $6^{(45)}$ ,   $\left( \tfrac{21}{4} \right)^{(20)}$ , $3^{(40)}$ , $\left( \tfrac{3}{4} \right)^{(64)}$ 
 \\[4pt]  
\cline{2-3} 
& $ 2$ & $20^{(2)}$ ,    $\left( \tfrac{65}{4} \right)^{(30)}$ , $14^{(60)}$ ,  $\left( \tfrac{53}{4} \right)^{(20)}$ , $11^{(60)}$ ,  $\left( \tfrac{35}{4} \right)^{(192)}$ ,  $8^{(90)}$ , $\left( \tfrac{17}{4} \right)^{(100)}$ ,  $2^{(175)}$
\\[4pt]  \hline \hline
\multirow{4}{*}{$\cN=0$, G$_2$} & $0$    & $3^{(14)}$ , $0^{(14)}$          \\[4pt]  
\cline{2-3} 
& $ 1$ & $12^{(2)}$ ,   $9^{(14)}$ ,   $6^{(28)}$ ,    $5^{(81)}$ , $\left( \tfrac{3}{2} \right)^{(64)}$ 
 \\[4pt]  
\cline{2-3} 
& $ 2$ & $17^{(21)}$ ,    $14^{(42)}$ ,   $13^{(54)}$ ,   $\left( \tfrac{19}{2} \right)^{(192)}$ , $8^{(231)}$ , $4^{(189)}$
\\[4pt]  \hline 
\end{tabular}


}\normalsize

\caption{\footnotesize{The KK vector spectra up to KK level $k=2$
about the massive type IIA AdS$_4$ solutions that uplift on $S^6$ from vacua of $D=4$ $\cN=8$ ISO(7) dyonically-gauged supergravity with at least SU(3) symmetry. The spectrum of the $\cN=3$ SO(4)-invariant solution is also included.}\normalsize}
\label{tab:KKVectorsISO7}
\end{table}

\end{landscape}


\section{$\cN=3$ supermultiplets} \label{sec:N=3Supermultiplets}

The state content of the supermultiplets of OSp$(4|2)$ has been conveniently tabulated in appendix A of \cite{Klebanov:2008vq}. These tables are very useful to allocate the spectrum of the $\cN=2$ AdS$_4$ solution discussed in section \ref{sec:N=2SpectrumIIA} into OSp$(4|2)$ supermultiplets. For OSp$(4|3)$, similar tables do not seem to be available in the literature: reference \cite{Fre:1999gok} tabulates the field content for representations of integer isospin only. In the spectrum of the $\cN=3$ AdS$_4$ solution of section \ref{sec:N=3SpectrumIIA}, multiplets of both integer and half-integer isospin arise. In this appendix, the state content of the OSp$(4|3)$ representations that appear in the spectrum of this solution are presented for convenience in tables \ref{MGRAV3j0=0}--\ref{SVEC3j0geq2} below.

The unitary representations of OSp$(4|3)$ are characterised by three quantum numbers: the Dynkin labels of the superconformal primary under the bosonic subalgebra $\textrm{SO}(3,2) \times \textrm{SO}(3)_\cR$. These are the $\textrm{SO}(3,2)$ spin $s_0$ and energy $E_0$, and the Dynkin label $j_0$ (usually referred to as isospin) of the R-symmetry group SO$(3)_\cR$. In the main text, I have denoted the isospin simply by $j$. Alternatively, the superconformal primary spin $s_0$ can be traded by the maximum $\textrm{SO}(3,2)$ spin $s_\textrm{max}$ present in the multiplet. This is the convention that I will use, thereby labelling OSp$(4|3)$ multiplets with $(s_\textrm{max} , E_0 , j_0)$. Using this convention, the multiplets can be given names according to the value of $s_\textrm{max}$: one can thus speak of graviton, gravitino, or vector multiplets, if $s_\textrm{max} =2$, $s_\textrm{max} =\tfrac32$, or $s_\textrm{max} =1$. There are no hypermultiplets (which would have $s_\textrm{max} =\tfrac12$). 

As for the range of the quantum numbers, in supergravity applications one only needs to consider the above three values of $s_\textrm{max}$. I use conventions in which the isospin is a non-negative half-integer: $j_0 =0 , \tfrac12 , 1 , \tfrac32 , 2 , \tfrac52 , 3 , \ldots$, so that the isospin $j_0$ representation of SO$(3)_\cR$ is $(2 j_0 +1)$-dimensional. Finally, $E_0$ is a real number subject to the unitarity bound $E_0 \geq j_0 + \tfrac32$ if $s_\textrm{max} = 2$, or $E_0 \geq j_0 +1$ if $s_\textrm{max} = \tfrac32$, or to the equality $E_0 = j_0$ if $s_\textrm{max} = 1$. In the first two cases, if the bound is saturated the multiplets are short (or massless), and long otherwise. Vector multiplets are always short (or massless). 

To summarise, there are seven types of OSp$(4|3)$ supermultiplets that may arise in supergravity applications:
\begin{enumerate}

\item \label{item:MGEV3} massless graviton multiplet (MGRAV$_3$), with $s_0 = \tfrac12$,  $j_0 =0$, $E_0 = j_0 +\tfrac32 = \tfrac32$;

\item short graviton multiplet (SGRAV$_3$), with $s_0 = \tfrac12$, $j_0 \geq \tfrac12$, $E_0 = j_0 +\tfrac32$;

\item long graviton multiplet (LGRAV$_3$), with $s_0 = \tfrac12$, $j_0 \geq  0 $, $E_0 > j_0 +\tfrac32$;

\item short gravitino multiplet (SGINO$_3$), with $s_0 = 0$, $j_0 \geq \tfrac12$, $E_0 = j_0 + 1$;

\item long gravitino multiplet (LGINO$_3$), with $s_0 = 0$, $j_0 \geq  0 $, $E_0 > j_0 + 1 $;

\item massless vector multiplet (MVEC$_3$), with $s_0 = 0$, $j_0 =1 $, $E_0 = j_0 =1$; and

\item \label{item:SVEC3}  short vector multiplet (SVEC$_3$), with $s_0 = 0$, $j_0 \geq \tfrac32 $, $E_0 = j_0$. 

\end{enumerate}
A massless gravitino multiplet exists, but it cannot arise in supergravity spectra as its presence would indicate an enhancement $\cN >3$ of supersymmetry. A subindex $3$ has been attached to the above acronyms in order to emphasise that they are $\cN=3$. 

The state content of these OSp$(4|3)$ supermultiplets is most conveniently described in terms of OSp$(4|2)$ supermultiplets. Denoting by MGRAV$_3 (E_0 , j_0)$, etc., the $\cN=3$ supermultiplets with energy $E_0$ and isospin $j_0$, and by MGRAV$_2 (E_0 , y_0)$, etc., the $\cN=2$ supermultiplets with energy $E_0$ and R-charge $y_0$, as given in appendix A of \cite{Klebanov:2008vq}, the OSp$(4|2)$ content of the OSp$(4|3)$ graviton multiplets is given by \cite{Fre:1999gok,Minwalla:2011ma}
{\setlength\arraycolsep{1pt}
\begin{eqnarray} \label{eqN=3multipGravitons}
\textrm{MGRAV}_3 \big( \tfrac32 , 0 \big) & =&  \textrm{MGRAV}_2 (2,0) \oplus  \textrm{MGINO}_2 (\tfrac12,0) \; , \nonumber \\[5pt]
\textrm{SGRAV}_3 \big( 2 , \tfrac12 \big) & =&  \textrm{SGRAV}_2 (\tfrac52 , -\tfrac12)  \oplus \textrm{SGRAV}_2 (\tfrac52 , \tfrac12)  \oplus \textrm{SGINO}_2 (2 , -\tfrac12)  \oplus \textrm{SGINO}_2 (2 , \tfrac12)  \; , \nonumber \\[5pt]
\textrm{SGRAV}_3 \big( E_0 , j_0 \big) & =& \textrm{SGRAV}_2 (E_0 +\tfrac12 , -j_0)  \oplus \textrm{SGRAV}_2 (E_0 +\tfrac12 , j_0)  \nonumber \\[3pt]
&& \oplus \bigoplus_{y_0 = -(j_0-1)}^{j_0-1}  \textrm{LGRAV}_2 (E_0 +\tfrac12 , y_0) \nonumber \\[3pt]
&& \oplus \, \textrm{SGINO}_2 (E_0 , -j_0) \oplus \textrm{SGINO}_2 (E_0 , j_0) \nonumber \\[3pt]
&& \oplus \bigoplus_{y_0 = -(j_0-1)}^{j_0-1}  \textrm{LGINO}_2 (E_0 , y_0)    \; , \quad \textrm{for $j_0 \geq 1$, $E_0 = j_0 + \tfrac32$}, \nonumber \\[5pt]
\textrm{LGRAV}_3 \big( E_0 , j_0 \big) & =&  
\bigoplus_{y_0 = -j_0}^{j_0}  \textrm{LGRAV}_2 (E_0 + \tfrac12 , y_0) \; 
\oplus  \bigoplus_{y_0 = -j_0}^{j_0}  \textrm{LGINO}_2 (E_0 +1 , y_0) \; \nonumber \\
&& \oplus  \bigoplus_{y_0 = -j_0}^{j_0}  \textrm{LGINO}_2 (E_0 , y_0)  \;  
\oplus  \bigoplus_{y_0 = -j_0}^{j_0}  \textrm{LVEC}_2 (E_0 + \tfrac12 , y_0) \; , \nonumber \\
&&  \textrm{for $j_0 \geq 0$, $E_0 > j_0 + \tfrac32$} \; .
\end{eqnarray}
}An $\cN=2$ massless gravitino multiplet (MGINO$_2$) was not tabulated in \cite{Klebanov:2008vq}, but its state content can be easily inferred. The OSp$(4|2)$ content of the OSp$(4|3)$ gravitino multiplets is given by \cite{Fre:1999gok,Minwalla:2011ma}
{\setlength\arraycolsep{1pt}
\begin{eqnarray} \label{eqN=3multipGravitinos}
\textrm{SGINO}_3 \big( \tfrac32, \tfrac12 \big) & =&  \textrm{SGINO}_2 (2 , -\tfrac12)  \oplus \textrm{SGINO}_2 (2 , \tfrac12)  \oplus \textrm{SVEC}_2 ( \tfrac32 , -\tfrac12)  \oplus \textrm{SVEC}_2 ( \tfrac32  , \tfrac12)  \; , \nonumber \\[5pt]
\textrm{SGINO}_3 \big( E_0 , j_0 \big) & =& \textrm{SGINO}_2 (E_0 +\tfrac12 , -j_0)  \oplus \textrm{SGINO}_2 (E_0 +\tfrac12 , j_0)  \nonumber \\[3pt]
&& \oplus \bigoplus_{y_0 = -(j_0-1)}^{j_0-1}  \textrm{LGINO}_2 (E_0 +\tfrac12 , y_0) \nonumber \\[3pt]
&& \oplus \, \textrm{SVEC}_2 (E_0 , -j_0) \oplus \textrm{SVEC}_2 (E_0 , j_0) \nonumber \\[3pt]
&& \oplus \bigoplus_{y_0 = -(j_0-1)}^{j_0-1}  \textrm{LVEC}_2 (E_0 , y_0)    \; , \quad \textrm{for $j_0 \geq \tfrac32$, $E_0 = j_0 + 1$}, \nonumber \\[5pt]
\textrm{LGINO}_3 \big( E_0 , j_0 \big) & =&  
\bigoplus_{y_0 = -j_0}^{j_0}  \textrm{LGINO}_2 (E_0 + \tfrac12 , y_0) \; 
\oplus  \bigoplus_{y_0 = -j_0}^{j_0}  \textrm{LVEC}_2 (E_0 +1 , y_0) \; \nonumber \\
&& \oplus  \bigoplus_{y_0 = -j_0}^{j_0}  \textrm{LVEC}_2 (E_0 , y_0) \; , \textrm{for $j_0 \geq 0$, $E_0 > j_0 + 1$} \; .
\end{eqnarray}
}Finally, the OSp$(4|2)$ content of the OSp$(4|3)$ vector multiplets is given by \cite{Fre:1999gok,Minwalla:2011ma}
{\setlength\arraycolsep{1pt}
\begin{eqnarray} \label{eqN=3multipVectors}
\textrm{MVEC}_3 \big( 1 , 1 \big) &=& \textrm{MVEC}_2 (1,0) \oplus  \textrm{HYP}_2 (1,-1) \oplus  \textrm{HYP}_2 (1,1) \; , \nonumber \\[5pt]
\textrm{SVEC}_3 \big( \tfrac32, \tfrac32 \big) & =&  \textrm{SVEC}_2 (\tfrac32 , -\tfrac12)  \oplus \textrm{SVEC}_2 ( \tfrac32 , \tfrac12)  \oplus \textrm{HYP}_2 ( \tfrac32 , -\tfrac32)  \oplus \textrm{HYP}_2 ( \tfrac32  , \tfrac32)  \; , \nonumber \\[5pt]
\textrm{SVEC}_3 \big( E_0 , j_0 \big) & =&  \textrm{SVEC}_2 (E_0 , -(j_0-1))  \oplus \textrm{SVEC}_2 ( E_0 , j_0-1) \nonumber \\
&& \oplus \bigoplus_{y_0 = -(j_0-2)}^{j_0-2}  \textrm{LVEC}_2 (E_0  , y_0) \; 
\oplus \textrm{HYP}_2 (E_0 , -j_0)\; 
\oplus \textrm{HYP}_2 (E_0 , j_0) \; , \nonumber \\[3pt]
&& \textrm{for $j_0 \geq 2$, $E_0 = j_0$} \; . 
\end{eqnarray}
}

The state content of the OSp$(4|3)$ supermultiplets in items \ref{item:MGEV3}--\ref{item:SVEC3} above can be worked out from (\ref{eqN=3multipGravitons})--(\ref{eqN=3multipVectors}), using the $\cN=2$ tables in appendix A of \cite{Klebanov:2008vq}. The result is presented in tables \ref{MGRAV3j0=0}--\ref{SVEC3j0geq2} below. These tables are useful to work out the $\cN=3$ spectrum discussed in section \ref{sec:N=3SpectrumIIA}.  Tables \ref{SGRAV3j0=1/2} and \ref{SGRAV3j0=3/2} for the SGRAV$_3$ with $j_0 = \frac12$ and $j_0 = \frac32$ are the only tables not needed for that purpose (according to table \ref{tab:ShortN=3Spectrum}, the short graviton spectrum for the $\cN=3$ AdS$_4$ solution comes in representations of integer isospin only), but are also presented for completeness and general reference. The tables for supermultiplets with integer isospin agree with \cite{Fre:1999gok} (except for table \ref{MVEC3j0=1} for the MVEC$_3$, which corrects a typo in their table 5 with $J_0 = 1$ therein). Unlike for $\cN=2$, where the state content of the OSp$(4|2)$ supermultiplets is independent of the R-charge $y_0$, in the $\cN=3$ case, supermultiplets of the same type ({\it e.g.} LGRAV$_3$'s) with different isospin $j_0$ may have different state contents.


\newpage

\begin{landscape}

\begin{center}



\begin{table}[H]
\begin{center}
{\scriptsize
\begin{tabular}{|l|c|c|c|c|}
\hline
spin	
& 	
$2$
&			
$\tfrac32$
&
$1$
&			
$\tfrac12$ \\[3pt]
\hline
\\[-10pt]
energy
& 	
$3$
&
$\tfrac52$
&
$2$
&
$\tfrac32$
\\[3pt]
\hline
\\[-10pt]
isospin
& 	
$0$
&
$1$  
&
$1$
&
0
\\
\hline
\end{tabular}
\caption{\footnotesize{$\cN=3$ massless graviton multiplet (MGRAV$_3$), $j_0=0$, $E_0 = j_0 +\tfrac32 = \tfrac32$ }\normalsize}
\label{MGRAV3j0=0}
}\normalsize
\end{center}
\end{table}



\begin{table}[H]
\begin{center}
{\scriptsize
\begin{tabular}{|l|c|c|c|c|c|c|c|c|}
\hline
spin	
& 	
$2$
&			
\multicolumn{2}{c|}{$\tfrac32$}
&
\multicolumn{2}{c|}{$1$} 
&
\multicolumn{2}{c|}{$\tfrac12$} 
&
$0$ \\
\hline
\\[-10pt]
energy
& 	
$\tfrac72$
&
$4$
&
$3$
&
$\tfrac72$
&
$\tfrac52$
&
$3$
&
$2$
&
$\tfrac52$
\\
\hline
\\[-9.5pt]
isospin
& 	
$\tfrac12$
&
$\tfrac12$ 
&
$\tfrac32$ , $\tfrac12$ 
&
$\tfrac32$ , $\tfrac12$ 
&
$\tfrac32$ , $\tfrac12$ 
&
$\tfrac32$ , $\tfrac12$ 
&
 $\tfrac12$ 
&
 $\tfrac12$ 
\\
\hline
\end{tabular}
\caption{\footnotesize{$\cN=3$ short graviton multiplet (SGRAV$_3$), $j_0 = \tfrac12 $, $E_0 = j_0+\tfrac32 = 2$}\normalsize}
\label{SGRAV3j0=1/2}
}\normalsize
\end{center}
\end{table}



\begin{table}[H]
\begin{center}
{\scriptsize
\begin{tabular}{|l|c|c|c|c|c|c|c|c|c|c|c|}
\hline
spin	
& 	
$2$
&			
\multicolumn{2}{c|}{$\tfrac32$}
&
\multicolumn{3}{c|}{$1$} 
&
\multicolumn{3}{c|}{$\tfrac12$} 
&
\multicolumn{2}{c|}{$0$} \\
\hline
\\[-10pt]
energy
& 	
$4$
&
$\tfrac92$
&
$\tfrac72$
&
$5$
&
$4$
&
$3$
&
$\tfrac92$
&
$\tfrac72$
&
$\tfrac52$
&
$4$
&
$3$
\\
\hline
\\[-9.5pt]
isospin
& 	
$1$
&
$1$ , $0$  
&
$2$ , $1$ , $0$ 
&
$0$  
&
$2$   , $1$   , $1$ , $0$ 
&
$2$   , $1$  , $0$ 
&
$1$
&
$2$   , $1$   , $1$ , $0$ 
&
$1$
&
$1$
&
$1$ , $0$
\\
\hline
\end{tabular}
\caption{\footnotesize{$\cN=3$ short graviton multiplet (SGRAV$_3$), $j_0 = 1$, $E_0 = j_0+\tfrac32 = \tfrac52$}\normalsize}
\label{SGRAV3j0=1}
}\normalsize
\end{center}
\end{table}

\newpage



\begin{table}[H]
\begin{center}
{\scriptsize
\begin{tabular}{|l|c|c|c|c|c|c|c|c|c|c|c|}
\hline
spin	
& 	
$2$
&			
\multicolumn{2}{c|}{$\tfrac32$}
&
\multicolumn{3}{c|}{$1$} 
&
\multicolumn{3}{c|}{$\tfrac12$} 
&
\multicolumn{2}{c|}{$0$} \\
\hline
\\[-10pt]
energy
& 	
$\tfrac92$
&
$5$
&
$4$
&
$\tfrac{11}{2}$
&
$\tfrac92$
&
$\tfrac72$
&
$5$
&
$4$
&
$3$
&
$\tfrac92$
&
$\tfrac72$
\\
\hline
\\[-9.5pt]
isospin
& 	
$\tfrac32$
&
$\tfrac32$ , $\tfrac12$
&
$\tfrac52$ , $\tfrac32$ , $\tfrac12$
&
$\tfrac12$
&
$\tfrac52$ , $\tfrac32$ , $\tfrac32$  , $\tfrac12$ , $\tfrac12$
&
$\tfrac52$ , $\tfrac32$ , $\tfrac12$
&
$\tfrac32$ , $\tfrac12$
&
$\tfrac52$ , $\tfrac32$ , $\tfrac32$  , $\tfrac12$ , $\tfrac12$
&
$\tfrac32$
&
$\tfrac32$ , $\tfrac12$
&
$\tfrac32$ , $\tfrac12$
\\
\hline
\end{tabular}
\caption{\footnotesize{$\cN=3$ short graviton multiplet (SGRAV$_3$), $j_0 = \tfrac32$, $E_0 = j_0+\tfrac32 = 3$}\normalsize}
\label{SGRAV3j0=3/2}
}\normalsize
\end{center}
\end{table}



\begin{table}[H]
\begin{center}
{\scriptsize

\begin{tabular}{|l|c|c|c|c|c|c|}
\hline
spin	
& 	
$2$
&			
\multicolumn{2}{c|}{$\tfrac32$}
&
\multicolumn{3}{c|}{$1$}  \\
\hline
\\[-10pt]
energy
& 	
$E_0 + \tfrac32$
&
$E_0 + 2$
&
$E_0 + 1$
&
$E_0 + \tfrac52$
&
$E_0 + \tfrac32$
&
$E_0 + \tfrac12$
\\
\hline
\\[-9.5pt]
isospin
& 	
$j_0$
&
$j_0$ , $j_0-1$  
&
$j_0+1$ , $j_0$ , $j_0-1$ 
&
$j_0-1$  
&
$j_0+1$   , $j_0$   , $j_0$ , $j_0-1$   , $j_0-1$   , $j_0-2$ 
&
$j_0+1 $ , $j_0$ ,  $j_0-1$
\\
\hline
\end{tabular}

\begin{tabular}{|l|c|c|c|c|c|}
\hline
spin	
& 	
\multicolumn{3}{c|}{$\tfrac12$} 
&
\multicolumn{2}{c|}{$0$} \\
\hline
\\[-10pt]
energy
&
$E_0 + 2$
&
$E_0 + 1$
&
$E_0$
&
$E_0 + \tfrac32$
&
$E_0 + \tfrac12$
\\
\hline
\\[-9.5pt]
isospin
&
$j_0$ , $j_0-1$ , $j_0-2$
&
$j_0+1$ , $j_0$ , $j_0$ , $j_0-1$   , $j_0-1$   , $j_0-2$ 
&
$j_0$
&
$j_0$ , $j_0-1$ , $j_0-2$
&
$j_0$ , $j_0-1$
\\
\hline
\end{tabular}

\caption{\footnotesize{$\cN=3$ short graviton multiplet (SGRAV$_3$), $j_0 \geq 2$, $E_0 = j_0+\tfrac32 $}\normalsize}
\label{SGRAV3j0geq2}
}\normalsize
\end{center}
\end{table}

\end{center}

\end{landscape}

\newpage

\begin{landscape}



\begin{table}[H]
\begin{center}
{\scriptsize
\begin{tabular}{|l|c|c|c|c|c|c|c|c|c|c|c|c|c|}
\hline
spin	
& 	
$2$
&			
\multicolumn{2}{c|}{$\tfrac32$}
&
\multicolumn{3}{c|}{$1$} 
&
\multicolumn{4}{c|}{$\tfrac12$} 
&
\multicolumn{3}{c|}{$0$} \\
\hline
\\[-10pt]
energy
& 	
$E_0 +\tfrac32$
&
$E_0 + 2$
&
$E_0 + 1$
&
$E_0 + \tfrac52$
&
$E_0 + \tfrac32$
&
$E_0 + \tfrac12$
&
$E_0 + 3$
&
$E_0 + 2$
&
$E_0 + 1$
&
$E_0 $
&
$E_0 + \tfrac52$
&
$E_0 + \tfrac32$
&
$E_0 + \tfrac12$
\\
\hline
\\[-9.5pt]
isospin
& 	
$0$
&
$1$  
&
$1$ 
&
$1$  
&
$2$   , $1$   , $0$ 
&
$1$ 
&
$0$
&
$2$  , $1$ , $0$ 
&
$2$  , $1$ , $0$ 
&
$0$
&
$1$
&
$2$ , $1$
&
$1$
\\
\hline
\end{tabular}
\caption{\footnotesize{$\cN=3$ long graviton multiplet (LGRAV$_3$), $j_0 = 0$, $E_0 > \tfrac32 $}\normalsize}
\label{LGRAV3j0=0}
}\normalsize
\end{center}
\end{table}


\begin{table}[H]
\begin{center}
{\scriptsize
\begin{tabular}{|l|c|c|c|c|c|c|c|c|c|c|c|c|c|}
\hline
spin	
& 	
$2$
&			
\multicolumn{2}{c|}{$\tfrac32$}
&
\multicolumn{3}{c|}{$1$} 
&
\multicolumn{4}{c|}{$\tfrac12$} 
&
\multicolumn{3}{c|}{$0$} \\
\hline
\\[-10pt]
energy
& 	
$E_0 +\tfrac32$
&
$E_0 + 2$
&
$E_0 + 1$
&
$E_0 + \tfrac52$
&
$E_0 + \tfrac32$
&
$E_0 + \tfrac12$
&
$E_0 + 3$
&
$E_0 + 2$
&
$E_0 + 1$
&
$E_0 $
&
$E_0 + \tfrac52$
&
$E_0 + \tfrac32$
&
$E_0 + \tfrac12$
\\
\hline
\\[-9.5pt]
isospin
& 	
$\tfrac12$
&
$\tfrac32$ , $\tfrac12$   
&
$\tfrac32$ , $\tfrac12$   
&
$\tfrac32$ , $\tfrac12$   
&
$\tfrac52$ , $\tfrac32$   , $\tfrac32$ , $\tfrac12$   , $\tfrac12$   
&
$\tfrac32$ , $\tfrac12$   
&
$\tfrac12$   
&
$\tfrac52$ , $\tfrac32$   , $\tfrac32$ , $\tfrac12$   , $\tfrac12$   
&
$\tfrac52$ , $\tfrac32$   , $\tfrac32$ , $\tfrac12$   , $\tfrac12$   
&
$\tfrac12$   
&
$\tfrac32$ , $\tfrac12$   
&
$\tfrac52$ , $\tfrac32$   , $\tfrac32$ , $\tfrac12$ 
&
$\tfrac32$ , $\tfrac12$   
\\
\hline
\end{tabular}
\caption{\footnotesize{$\cN=3$ long graviton multiplet (LGRAV$_3$), $j_0 = \tfrac12$, $E_0 > 2 $}\normalsize}
\label{LGRAV3j0=1/2}
}\normalsize
\end{center}
\end{table}



\begin{table}[H]
\begin{center}
{\scriptsize

\begin{tabular}{|l|c|c|c|c|c|c|}
\hline
spin	
& 	
$2$
&			
\multicolumn{2}{c|}{$\tfrac32$}
&
\multicolumn{3}{c|}{$1$}  \\
\hline
\\[-10pt]
energy
& 	
$E_0 + \tfrac32$
&
$E_0 + 2$
&
$E_0 + 1$
&
$E_0 + \tfrac52$
&
$E_0 + \tfrac32$
&
$E_0 + \tfrac12$
\\
\hline
\\[-9.5pt]
isospin
& 	
$1$
&
$2$ , $1$ , $0$  
&
$2$ , $1$ , $0$
&
$2$ , $1$ , $0$
&
$3$ , $2$ , $2$ , $1$ , $1$ , $1$ , $0$ 
&
$2$ , $1$ , $0$
\\
\hline
\end{tabular}

\begin{tabular}{|l|c|c|c|c|c|c|c|}
\hline
spin	
& 	
\multicolumn{4}{c|}{$\tfrac12$} 
&
\multicolumn{3}{c|}{$0$} \\
\hline
\\[-10pt]
energy
&
$E_0 + 3$
&
$E_0 + 2$
&
$E_0 + 1$
&
$E_0$
&
$E_0 + \tfrac52$
&
$E_0 + \tfrac32$
&
$E_0 + \tfrac12$
\\
\hline
\\[-9.5pt]
isospin
&
$1$
&
$3$ , $2$ , $2$ , $1$ , $1$ , $1$ , $0$ 
&
$3$ , $2$ , $2$ , $1$ , $1$ , $1$ , $0$ 
&
$1$
&
$2$ , $1$ , $0$
&
$3$ , $2$ , $2$ , $1$ , $1$ , $0$ 
&
$2$ , $1$ , $0$
\\
\hline
\end{tabular}

\caption{\footnotesize{$\cN=3$ long graviton multiplet (LGRAV$_3$), $j_0 =1$, $E_0 > \tfrac52 $}\normalsize}
\label{LGRAV3j0=1}
}\normalsize
\end{center}
\end{table}

\newpage


\begin{table}[H]
\begin{center}
{\scriptsize

\begin{tabular}{|l|c|c|c|c|c|c|}
\hline
spin	
& 	
$2$
&			
\multicolumn{2}{c|}{$\tfrac32$}
&
\multicolumn{3}{c|}{$1$}  \\
\hline
\\[-10pt]
energy
& 	
$E_0 + \tfrac32$
&
$E_0 + 2$
&
$E_0 + 1$
&
$E_0 + \tfrac52$
&
$E_0 + \tfrac32$
&
$E_0 + \tfrac12$
\\
\hline
\\[-9.5pt]
isospin
& 	
$\tfrac32$
&
$\tfrac52$ , $\tfrac32$ , $\tfrac12$
&
$\tfrac52$ , $\tfrac32$ , $\tfrac12$
&
$\tfrac52$ , $\tfrac32$ , $\tfrac12$
&
$\tfrac72$ , $\tfrac52$ , $\tfrac52$ , $\tfrac32$ , $\tfrac32$ , $\tfrac32$ , $\tfrac12$ , $\tfrac12$
&
$\tfrac52$ , $\tfrac32$ , $\tfrac12$
\\
\hline
\end{tabular}

\begin{tabular}{|l|c|c|c|c|c|c|c|}
\hline
spin	
& 	
\multicolumn{4}{c|}{$\tfrac12$} 
&
\multicolumn{3}{c|}{$0$} \\
\hline
\\[-10pt]
energy
&
$E_0 + 3$
&
$E_0 + 2$
&
$E_0 + 1$
&
$E_0$
&
$E_0 + \tfrac52$
&
$E_0 + \tfrac32$
&
$E_0 + \tfrac12$
\\
\hline
\\[-9.5pt]
isospin
&
$\tfrac32$
&
$\tfrac72$ , $\tfrac52$ , $\tfrac52$ , $\tfrac32$ , $\tfrac32$ , $\tfrac32$ , $\tfrac12$ , $\tfrac12$
&
$\tfrac72$ , $\tfrac52$ , $\tfrac52$ , $\tfrac32$ , $\tfrac32$ , $\tfrac32$ , $\tfrac12$ , $\tfrac12$
&
$\tfrac32$
&
$\tfrac52$ , $\tfrac32$ , $\tfrac12$ 
&
$\tfrac72$ , $\tfrac52$ , $\tfrac52$ , $\tfrac32$ , $\tfrac32$ ,  $\tfrac12$ , $\tfrac12$
&
$\tfrac52$ , $\tfrac32$ , $\tfrac12$ 
\\
\hline
\end{tabular}

\caption{\footnotesize{$\cN=3$ long graviton multiplet (LGRAV$_3$), $j_0 =\tfrac32$, $E_0 > 3 $}\normalsize}
\label{LGRAV3j0=3/2}
}\normalsize
\end{center}
\end{table}



\begin{table}[H]
\begin{center}
{\scriptsize

\begin{tabular}{|l|c|c|c|c|c|c|}
\hline
spin	
& 	
$2$
&			
\multicolumn{2}{c|}{$\tfrac32$}
&
\multicolumn{3}{c|}{$1$}  \\
\hline
\\[-10pt]
energy
& 	
$E_0 + \tfrac32$
&
$E_0 + 2$
&
$E_0 + 1$
&
$E_0 + \tfrac52$
&
$E_0 + \tfrac32$
&
$E_0 + \tfrac12$
\\
\hline
\\[-9.5pt]
isospin
& 	
$j_0$
&
$j_0+1$ , $j_0$ , $j_0-1$  
&
$j_0+1$ , $j_0$ , $j_0-1$ 
&
$j_0+1$ , $j_0$ , $j_0-1$ 
&
$j_0+2$ , $j_0+1$ , $j_0+1$ , $j_0$ , $j_0$ , $j_0$ , $j_0-1$, $j_0-1$, $j_0-2$ 
&
$j_0+1$ , $j_0$ , $j_0-1$ 
\\
\hline
\end{tabular}

\begin{tabular}{|l|c|c|c|c|}
\hline
spin	
& 	
\multicolumn{4}{c|}{$\tfrac12$}  \\
\hline
\\[-10pt]
energy
&
$E_0 + 3$
&
$E_0 + 2$
&
$E_0 + 1$
&
$E_0$
\\
\hline
\\[-9.5pt]
isospin
&
$j_0$
&
$j_0+2$ , $j_0+1$ , $j_0+1$ , $j_0$ , $j_0$ , $j_0$ , $j_0-1$, $j_0-1$, $j_0-2$ 
&
$j_0+2$ , $j_0+1$ , $j_0+1$ , $j_0$ , $j_0$ , $j_0$ , $j_0-1$, $j_0-1$, $j_0-2$ 
&
$j_0$
\\
\hline
\end{tabular}

\begin{tabular}{|l|c|c|c|}
\hline
spin	
&
\multicolumn{3}{c|}{$0$} \\
\hline
\\[-10pt]
energy
&
$E_0 + \tfrac52$
&
$E_0 + \tfrac32$
&
$E_0 + \tfrac12$
\\
\hline
\\[-9.5pt]
isospin
&
 $j_0+1$ ,  $j_0$ ,  $j_0-1$ 
&
$j_0+2$ , $j_0+1$ , $j_0+1$ , $j_0$ , $j_0$ ,  $j_0-1$, $j_0-1$, $j_0-2$ 
&
 $j_0+1$ ,  $j_0$ ,  $j_0-1$ 
\\
\hline
\end{tabular}

\caption{\footnotesize{$\cN=3$ long graviton multiplet (LGRAV$_3$), $j_0 \geq 2$, $E_0 > j_0 + \tfrac32$}}\normalsize
\label{LGRAV3j0geq2}
}\normalsize
\end{center}
\end{table}

\end{landscape}

\newpage

\begin{landscape}





\begin{table}[H]
\begin{center}
{\scriptsize
\begin{tabular}{|l|c|c|c|c|c|c|c|}
\hline
spin	
& 	
$\tfrac32$
&			
\multicolumn{2}{c|}{$1$}
&
\multicolumn{2}{c|}{$\tfrac12$} 
&
\multicolumn{2}{c|}{$0$} \\
\hline
\\[-10pt]
energy
& 	
$3$
&
$\tfrac72$
&
$\tfrac52$
&
$3$
&
$2$
&
$\tfrac52$
&
$\tfrac32$
\\
\hline
\\[-9.5pt]
isospin
& 	
$\tfrac12$
&
$\tfrac12$ 
&
$\tfrac32$ , $\tfrac12$ 
&
$\tfrac32$ , $\tfrac12$ 
&
$\tfrac32$ , $\tfrac12$ 
&
$\tfrac32$ , $\tfrac12$ 
&
 $\tfrac12$ 
\\
\hline
\end{tabular}
\caption{\footnotesize{$\cN=3$ short gravitino multiplet (SGINO$_3$), $j_0 = \tfrac12$, $E_0 = j_0+1 = \tfrac32$}}\normalsize
\label{SGINO3j0=1/2}
}\normalsize
\end{center}
\end{table}




\begin{table}[H]
\begin{center}
{\scriptsize
\begin{tabular}{|l|c|c|c|c|c|c|c|c|c|}
\hline
spin	
& 	
$\tfrac32$
&			
\multicolumn{2}{c|}{$1$}
&
\multicolumn{3}{c|}{$\tfrac12$} 
&
\multicolumn{3}{c|}{$0$} \\
\hline
\\[-10pt]
energy
& 	
$\tfrac72$
&
$4$
&
$3$
&
$\tfrac92$
&
$\tfrac72$
&
$\tfrac52$
&
$4$
&
$3$
&
$2$
\\
\hline
\\[-9.5pt]
isospin
& 	
$1$
&
$1$ , $0$  
&
$2$ , $1$ , $0$ 
&
$0$
&
$2$ , $1$ , $1$ , $0$ 
&
$2$ , $1$ , $0$ 
&
$1$
&
$2$ , $1$ , $1$ 
&
$1$  
\\
\hline
\end{tabular}
\caption{\footnotesize{$\cN=3$ short gravitino multiplet (SGINO$_3$), $j_0 = 1$, $E_0 = j_0+1 = 2$}}\normalsize
\label{SGINO3j0=1}
}\normalsize
\end{center}
\end{table}




\begin{table}[H]
\begin{center}
{\scriptsize
\begin{tabular}{|l|c|c|c|c|c|c|c|c|c|}
\hline
spin	
& 	
$\tfrac32$
&			
\multicolumn{2}{c|}{$1$}
&
\multicolumn{3}{c|}{$\tfrac12$} 
&
\multicolumn{3}{c|}{$0$} \\
\hline
\\[-10pt]
energy
& 	
$4$
&
$\tfrac92$
&
$\tfrac72$
&
$5$
&
$4$
&
$3$
&
$\tfrac92$
&
$\tfrac72$
&
$\tfrac52$
\\
\hline
\\[-9.5pt]
isospin
& 	
$\tfrac32$
&
$\tfrac32$ , $\tfrac12$  
&
$\tfrac52$ , $\tfrac32$ , $\tfrac12$ 
&
$\tfrac12$  
&
$\tfrac52$   , $\tfrac32$   , $\tfrac32$ , $\tfrac32$ , $\tfrac12$ 
&
$\tfrac52$ , $\tfrac12$ , $\tfrac12$ 
&
$\tfrac32$ , $\tfrac12$
&
$\tfrac52$ , $\tfrac32$ , $\tfrac32$ , $\tfrac12$
&
$\tfrac32$
\\
\hline
\end{tabular}
\caption{\footnotesize{$\cN=3$ short gravitino multiplet (SGINO$_3$), $j_0 = \tfrac32$, $E_0 = j_0+1 = \tfrac52$}\normalsize}
\label{SGINO3j0=3/2}
}\normalsize
\end{center}
\end{table}



\begin{table}[H]
\begin{center}
{\scriptsize
\begin{tabular}{|l|c|c|c|c|c|c|c|c|c|}
\hline
spin 
& 	
$\tfrac32$
&			
\multicolumn{2}{c|}{$1$}
&
\multicolumn{3}{c|}{$\tfrac12$}  \\
\hline
\\[-10pt]
energy
& 	
$E_0+\tfrac32$
&
$E_0+2$
&
$E_0+1$
&
$E_0+\tfrac52$
&
$E_0+\tfrac32$
&
$E_0+\tfrac12$
\\
\hline
\\[-9.5pt]
isospin
& 	
$j_0$
&
$j_0$ , $j_0-1$  
&
$j_0+1$ , $j_0$ , $j_0-1$ 
&
$j_0-1$
&
$j_0+1$ , $j_0$ , $j_0$ , $j_0-1$  , $j_0-1$  , $j_0-2$ 
&
$j_0+1$ , $j_0$ , $j_0-1$ 
\\
\hline
\end{tabular}

\begin{tabular}{|l|c|c|c|c|c|c|c|c|c|}
\hline
spin 
&
\multicolumn{3}{c|}{$0$} \\
\hline
\\[-10pt]
energy
&
$E_0+2$
&
$E_0+1$
&
$E_0$
\\
\hline
\\[-9.5pt]
isospin
&
$j_0$ , $j_0-1$ , $j_0-2$
&
$j_0+1$ , $j_0$ , $j_0$ , $j_0-1$, $j_0-2$
&
$j_0$
\\
\hline
\end{tabular}

\caption{\footnotesize{$\cN=3$ short gravitino multiplet (SGINO$_3$),  $j_0 \geq 2$, $E_0 = j_0+1$}\normalsize}
\label{SGINO3j0geq2}
}\normalsize
\end{center}
\end{table}

\end{landscape}

\newpage

\begin{landscape}



\vspace{-50pt}

\begin{table}[H]
\begin{center}
{\scriptsize
\begin{tabular}{|l|c|c|c|c|c|c|c|c|c|c|}
\hline
spin	
& 	
$\tfrac32$
&			
\multicolumn{2}{c|}{$1$}
&
\multicolumn{3}{c|}{$\tfrac12$} 
&
\multicolumn{4}{c|}{$0$} \\
\hline
\\[-10pt]
energy
& 	
$E_0 + \tfrac32$
&
$E_0 + 2$
&
$E_0 + 1$
&
$E_0 + \tfrac52$
&
$E_0 + \tfrac32$
&
$E_0 + \tfrac12$
&
$E_0 + 3$
&
$E_0 + 2$
&
$E_0 + 1$
&
$E_0 $
\\
\hline
\\[-9.5pt]
isospin
& 	
$0$
&
$1$ 
&
$1$ 
&
 $1$  
&
$2$ , $1$  
&
 $1$  
&
$0$  
&
$2$ , $0$ 
&
$2$ , $0$ 
&
$0$  
\\
\hline
\end{tabular}
\caption{\footnotesize{$\cN=3$ long gravitino multiplet (LGINO$_3$), $j_0 = 0$, $E_0 > 1$}\normalsize}
\label{LGINO3j0=0}
}\normalsize
\end{center}
\end{table}

\vspace{-10pt}


\begin{table}[H]
\begin{center}
{\scriptsize
\begin{tabular}{|l|c|c|c|c|c|c|c|c|c|c|}
\hline
spin	
& 	
$\tfrac32$
&			
\multicolumn{2}{c|}{$1$}
&
\multicolumn{3}{c|}{$\tfrac12$} 
&
\multicolumn{4}{c|}{$0$} \\
\hline
\\[-10pt]
energy
& 	
$E_0 + \tfrac32$
&
$E_0 + 2$
&
$E_0 + 1$
&
$E_0 + \tfrac52$
&
$E_0 + \tfrac32$
&
$E_0 + \tfrac12$
&
$E_0 + 3$
&
$E_0 + 2$
&
$E_0 + 1$
&
$E_0 $
\\
\hline
\\[-9.5pt]
isospin
& 	
$\tfrac12$
&
$\tfrac32$ , $\tfrac12$  
&
$\tfrac32$ , $\tfrac12$  
&
$\tfrac32$ , $\tfrac12$  
&
$\tfrac52$ , $\tfrac32$ , $\tfrac32$ , $\tfrac12$  
&
$\tfrac32$ , $\tfrac12$  
&
$\tfrac12$  
&
$\tfrac52$ , $\tfrac32$  , $\tfrac12$  
&
$\tfrac52$ , $\tfrac32$  , $\tfrac12$  
&
$\tfrac12$  
\\
\hline
\end{tabular}
\caption{\footnotesize{$\cN=3$ long gravitino multiplet (LGINO$_3$), $j_0 = \tfrac{1}{2}$, $E_0 > \tfrac32$}\normalsize}
\label{LGINO3j0=1/2}
}\normalsize
\end{center}
\end{table}



\vspace{-10pt}


\begin{table}[H]
\begin{center}
{\scriptsize
\begin{tabular}{|l|c|c|c|c|c|c|c|c|c|c|}
\hline
spin	
& 	
$\tfrac32$
&			
\multicolumn{2}{c|}{$1$}
&
\multicolumn{3}{c|}{$\tfrac12$} 
&
\multicolumn{4}{c|}{$0$} \\
\hline
\\[-10pt]
energy
& 	
$E_0 + \tfrac32$
&
$E_0 + 2$
&
$E_0 + 1$
&
$E_0 + \tfrac52$
&
$E_0 + \tfrac32$
&
$E_0 + \tfrac12$
&
$E_0 + 3$
&
$E_0 + 2$
&
$E_0 + 1$
&
$E_0 $
\\
\hline
\\[-9.5pt]
isospin
& 	
$1$
&
$2$ , $1$  , $0$
&
$2$ , $1$  , $0$
&
$2$ , $1$  , $0$
&
$3$ , $2$  , $2$ , $1$ , $1$  , $0$
&
$2$ , $1$  , $0$
&
$1$  
&
$3$ , $2$  , $1$ , $1$ 
&
$3$ , $2$  , $1$ , $1$ 
&
$1$  
\\
\hline
\end{tabular}
\caption{\footnotesize{$\cN=3$ long gravitino multiplet (LGINO$_3$), $j_0 = 1$, $E_0 > 2$}\normalsize}
\label{LGINO3j0=1}
}\normalsize
\end{center}
\end{table}


\vspace{-10pt}


\begin{table}[H]
\begin{center}
{\scriptsize
\begin{tabular}{|l|c|c|c|c|c|c|c|c|c|c|}
\hline
spin	
& 	
$\tfrac32$
&			
\multicolumn{2}{c|}{$1$}
&
\multicolumn{3}{c|}{$\tfrac12$} 
&
\multicolumn{4}{c|}{$0$} \\
\hline
\\[-10pt]
energy
& 	
$E_0 + \tfrac32$
&
$E_0 + 2$
&
$E_0 + 1$
&
$E_0 + \tfrac52$
&
$E_0 + \tfrac32$
&
$E_0 + \tfrac12$
&
$E_0 + 3$
&
$E_0 + 2$
&
$E_0 + 1$
&
$E_0 $
\\
\hline
\\[-9.5pt]
isospin
& 	
$\tfrac32$
&
$\tfrac52$ , $\tfrac32$  , $\tfrac12$
&
$\tfrac52$ , $\tfrac32$  , $\tfrac12$
&
$\tfrac52$ , $\tfrac32$  , $\tfrac12$
&
$\tfrac72$ , $\tfrac52$  , $\tfrac52$ , $\tfrac32$ , $\tfrac32$  , $\tfrac12$ , $\tfrac12$
&
$\tfrac52$ , $\tfrac32$  , $\tfrac12$
&
$\tfrac32$
&
$\tfrac72$ , $\tfrac52$  , $\tfrac32$ , $\tfrac32$ , $\tfrac12$ 
&
$\tfrac72$ , $\tfrac52$  , $\tfrac32$ , $\tfrac32$ , $\tfrac12$ 
&
$\tfrac32$ 
\\
\hline
\end{tabular}
\caption{\footnotesize{$\cN=3$ long gravitino multiplet (LGINO$_3$), $j_0 = \tfrac32$, $E_0 > \tfrac52$}\normalsize}
\label{LGINO3j0=3/2}
}\normalsize
\end{center}
\end{table}


\vspace{-10pt}


\begin{table}[H]
\begin{center}
{\scriptsize

\begin{tabular}{|l|c|c|c|c|c|c|}
\hline
spin	
& 	
$\tfrac32$
&			
\multicolumn{2}{c|}{$1$}
&
\multicolumn{3}{c|}{$\tfrac12$}  \\
\hline
\\[-10pt]
energy
& 	
$E_0 + \tfrac32$
&
$E_0 + 2$
&
$E_0 + 1$
&
$E_0 + \tfrac52$
&
$E_0 + \tfrac32$
&
$E_0 + \tfrac12$
\\
\hline
\\[-9.5pt]
isospin
& 	
$j_0$
&
$j_0 +1 $ , $j_0$  , $j_0 -1$
&
$j_0 +1 $ , $j_0$  , $j_0 -1$
&
$j_0 +1 $ , $j_0$  , $j_0 -1$
&
$j_0 +2 $ , $j_0 +1$  , $j_0 +1$ , $j_0 $ , $j_0$  , $j_0 -1$  , $j_0 -1$  , $j_0 -2$ 
&
$j_0 +1 $ , $j_0$  , $j_0 -1$
\\
\hline
\end{tabular}

\begin{tabular}{|l|c|c|c|c|c|c|c|c|c|c|}
\hline
spin	
&
\multicolumn{4}{c|}{$0$} \\
\hline
\\[-10pt]
energy
&
$E_0 + 3$
&
$E_0 + 2$
&
$E_0 + 1$
&
$E_0 $
\\
\hline
\\[-9.5pt]
isospin
&
$j_0  $
&
$j_0 +2 $ , $j_0 +1$  ,  $j_0 $ , $j_0$  , $j_0 -1$  ,  $j_0 -2$ 
&
$j_0 +2 $ , $j_0 +1$  ,  $j_0 $ , $j_0$  , $j_0 -1$  ,  $j_0 -2$ 
&
$j_0  $
\\
\hline
\end{tabular}

\caption{\footnotesize{$\cN=3$ long gravitino multiplet (LGINO$_3$), $j_0 \geq 2$, $E_0 > j_0 +1 $}\normalsize}
\label{LGINO3j0geq2}
}\normalsize
\end{center}
\end{table}


\end{landscape}

\newpage 

\begin{landscape}


\begin{table}[H]
\begin{center}
{\scriptsize
\begin{tabular}{|l|c|c|c|c|}
\hline
spin	
& 	
$1$
&			
$\tfrac12$
&
\multicolumn{2}{c|}{$0$} \\
\hline
\\[-10pt]
energy
& 	
$2$
&
$\tfrac32$
&
$2$
&
$1$
\\
\hline
\\[-9.5pt]
isospin
& 	
$0$
&
$1$ , $0$ 
&
$1$ 
&
$1$ 
\\
\hline
\end{tabular}
\caption{\footnotesize{$\cN=3$ massless vector multiplet (MVEC$_3$), $j_0 = 1$, $E_0 = j_0 = 1 $}}\normalsize
\label{MVEC3j0=1}
}\normalsize
\end{center}
\end{table}


\begin{table}[H]
\begin{center}
{\scriptsize
\begin{tabular}{|l|c|c|c|c|c|}
\hline
spin	
& 	
$1$
&			
\multicolumn{2}{c|}{$\tfrac12$} 
&
\multicolumn{2}{c|}{$0$} \\
\hline
\\[-10pt]
energy
& 	
$\tfrac52$
&
$3$
&
$2$
&
$\tfrac52$
&
$\tfrac32$
\\
\hline
\\[-9.5pt]
isospin
& 	
$\tfrac12$
&
$\tfrac12$
&
$\tfrac32$ , $\tfrac12$
&
$\tfrac32$ , $\tfrac12$
&
$\tfrac32$ 
\\
\hline
\end{tabular}
\caption{\footnotesize{$\cN=3$ short vector multiplet (SVEC$_3$), $j_0 = \tfrac32$, $E_0 = j_0 = \tfrac32 $}}\normalsize
\label{SVEC3j0=3/2}
}\normalsize
\end{center}
\end{table}


\begin{table}[H]
\begin{center}
{\scriptsize
\begin{tabular}{|l|c|c|c|c|c|c|}
\hline
spin	
& 	
$1$
&			
\multicolumn{2}{c|}{$\tfrac12$} 
&
\multicolumn{3}{c|}{$0$} \\
\hline
\\[-10pt]
energy
& 	
$E_0 +1 $
&
$E_0 + \tfrac32$
&
$E_0 + \tfrac12$
&
$E_0 + 2$
&
$E_0 + 1$
&
$E_0 $
\\
\hline
\\[-9.5pt]
isospin
& 	
$j_0-1$
&
$j_0-1$ , $j_0-2$
&
$j_0$ , $j_0-1$
&
$j_0-2$
&
$j_0$ , $j_0-1$ , $j_0-2$
&
$j_0$
\\
\hline
\end{tabular}
\caption{\footnotesize{$\cN=3$ short vector multiplet (SVEC$_3$), $j_0 \geq 2$, $E_0 =  j_0 $}}\normalsize
\label{SVEC3j0geq2}
}\normalsize
\end{center}
\end{table}


\end{landscape}

\bibliography{references}

\end{document}